\documentclass[final,3p,times]{elsarticle}

\usepackage{amssymb}
\usepackage{graphicx} 
\usepackage{pdflscape}
\usepackage{longtable}
\usepackage{array}
\usepackage{booktabs}
\usepackage{geometry}
\usepackage{tabularray}

\usepackage{array, booktabs, xcolor, siunitx}
\usepackage{caption}
\usepackage{multirow}

\usepackage{xcolor}
\usepackage{changes}
\usepackage{ulem}
\usepackage{xcolor}
\usepackage{rotating}
\usepackage{capt-of}
\usepackage{hyperref}
\usepackage{float}
\usepackage{array}
\usepackage{tikz}
\usepackage{longtable}
\usepackage{tabularx}
\usepackage{pdflscape}
\usepackage{amsmath}
\usepackage{enumitem}
\usepackage{hyperref}
\usepackage{subcaption}

\usepackage{amsmath}
\usepackage[table]{xcolor}
\usepackage{amsmath}
\usepackage{tikz}
\usetikzlibrary{tikzmark}
\usetikzlibrary{arrows.meta}

\usepackage{amssymb}
\usepackage{pifont}
\renewcommand{\arraystretch}{1.5}
\journal{Computer \& Security}

\begin{document}

\begin{frontmatter}

\title{Privacy Engineering in Smart Home (SH) Systems: A Comprehensive Privacy Threat Analysis (PTA) and Risk Management Approach}

\author{Emmanuel Dare Alalade} 

\affiliation{organization={School of Information Technology},
            addressline={Carleton University}, 
            city={Ottawa},
            postcode={K1S 5B6}, 
            state={Ontario},
            country={Canada}}
\author{Mohammed Mayhoub} 

\author{Ashraf Matrawy} 

\begin{abstract}
Addressing trust concerns in Smart Home (SH) systems is imperative due to the limited study on privacy preservation approaches that focus on analyzing and evaluating privacy threats for effective risk management. While most research focuses primarily on user privacy,  device data privacy, especially device data that can reveal the identity of a smart device, is almost neglected. Moreover, a breach of device data could indirectly reveal the user data, as these smart devices store, process, and transmit user and device data within the SH system. To this end, our study incorporates Privacy Engineering (PE) principles in the SH system that considers user and device data privacy, leveraging the LINDDUN PRO Privacy Engineering (PE) framework. In the first stage of the LINDDUN PRO PE framework, we present a DFD based on a typical SH reference model to better understand SH system operations. To identify potential areas vulnerable to privacy threats in the SH system, we conduct a Privacy Threat Analysis (PTA) using the LINDDUN PRO threat model in the second stage. In the third stage, a Privacy Impact Assessment (PIA) was conducted to manage privacy risks by prioritizing privacy threats based on their likelihood of occurrence and potential consequences. Finally, we suggest possible Privacy-Enhancing Technologies (PETs) that can mitigate some of these privacy threats. Furthermore, we validate the impact of implementing PETs to minimize the privacy risks in the SH system. This study aims to elucidate the primary privacy threats in SH systems, their associated risks, and the prioritization of these threats in SH systems, as well as the effect of PETs implementation on the SH system. The outcomes of this study are expected to benefit SH stakeholders, including vendors, cloud providers, users, researchers, and regulatory bodies in the SH systems domain.  
\end{abstract}


\end{frontmatter}

\section{Introduction}
\label{introduction}
 Smart Home (SH) has brought technical advancement to household operations by automating the interaction of home devices for convenience, comfort, security, and entertainment \cite{alalade2020intrusion}. However, in addition to this benefit, the SH system also poses some privacy concerns due to vulnerabilities associated with smart devices, communication protocols, channels, and firmware that make up a SH system \cite{HAMMI2022102677}. Exploiting these vulnerabilities by threat actors could undermine user benefits and trust, potentially leading to privacy breaches and distrust among users in the SH system.

 While there have been significant studies regarding vulnerabilities exploitable by privacy threats and different privacy measures to mitigate these privacy threats, most of these studies focus on analyzing and assessing privacy threats and their risks on user data within a SH \cite{zhao2020privacy, jose2016improving, shin2017secure, xu2019privacy, nimmy2021lightweight}. However, there have been limited studies on privacy preservation of smart device data that could reveal their identity \cite{abi2018preserving}. A breach in an Internet of Things (IoT) system, such as a Smart Home (SH), can compromise the identities of both smart devices and their users by exposing sensitive information collected and stored within these devices \cite{zohaib2023automated, mustafa2021iot}. For instance, an attacker may intercept smart device communication data to extract sensitive data such as the device’s MAC address, thereby revealing its unique identity \cite{lu2020nowhere}. Such exposure not only endangers the smart device itself but also indirectly discloses the identity of the SH user, either through their interactions with the smart device or through personal data stored on it. Given the importance of smart devices in the SH system, protecting their identities is crucial, as they store, process, and transmit user and smart device data within the SH system \cite{van2019internet}.

\begin{figure}
        \centering
        \includegraphics[width=10cm,height=6cm]{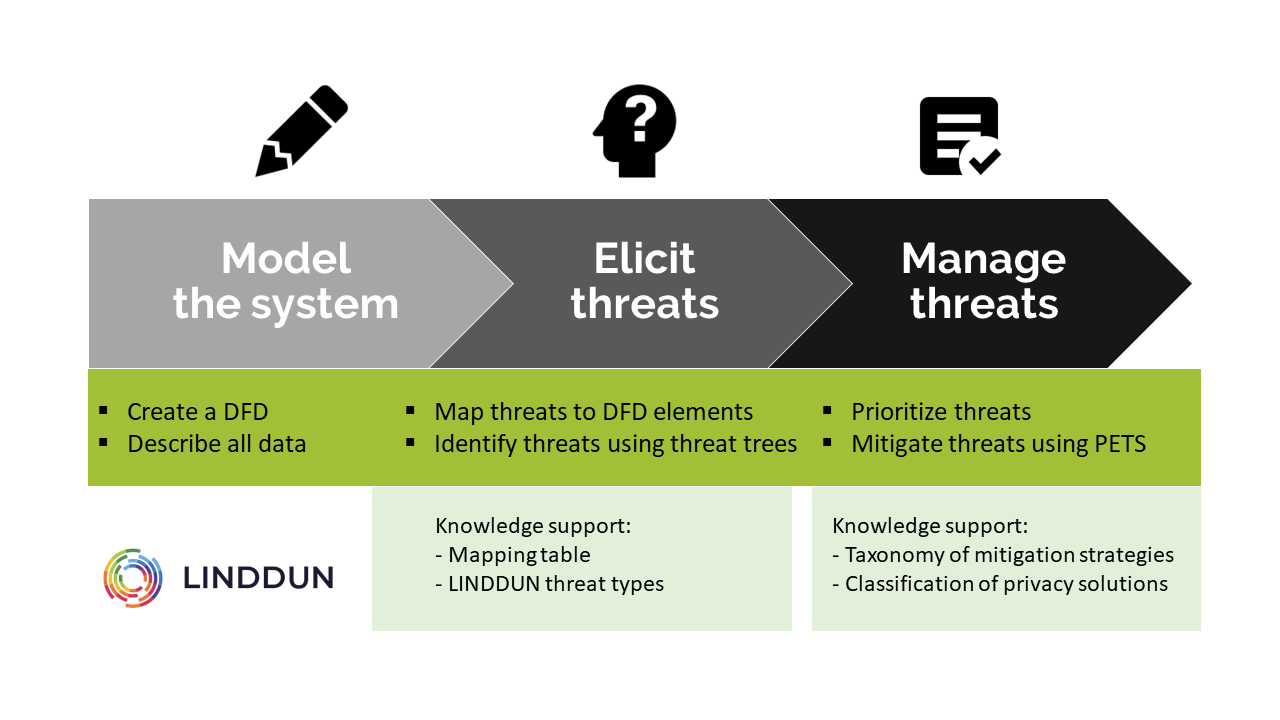}
        \caption{LINDDUN PRO PE Framework \cite{LINDDUN}}
        \label{fig:LINDDUN-framework}
\end{figure}

\begin{figure}[ht]
        \centering
        \includegraphics[width=9.5cm,height=7cm]{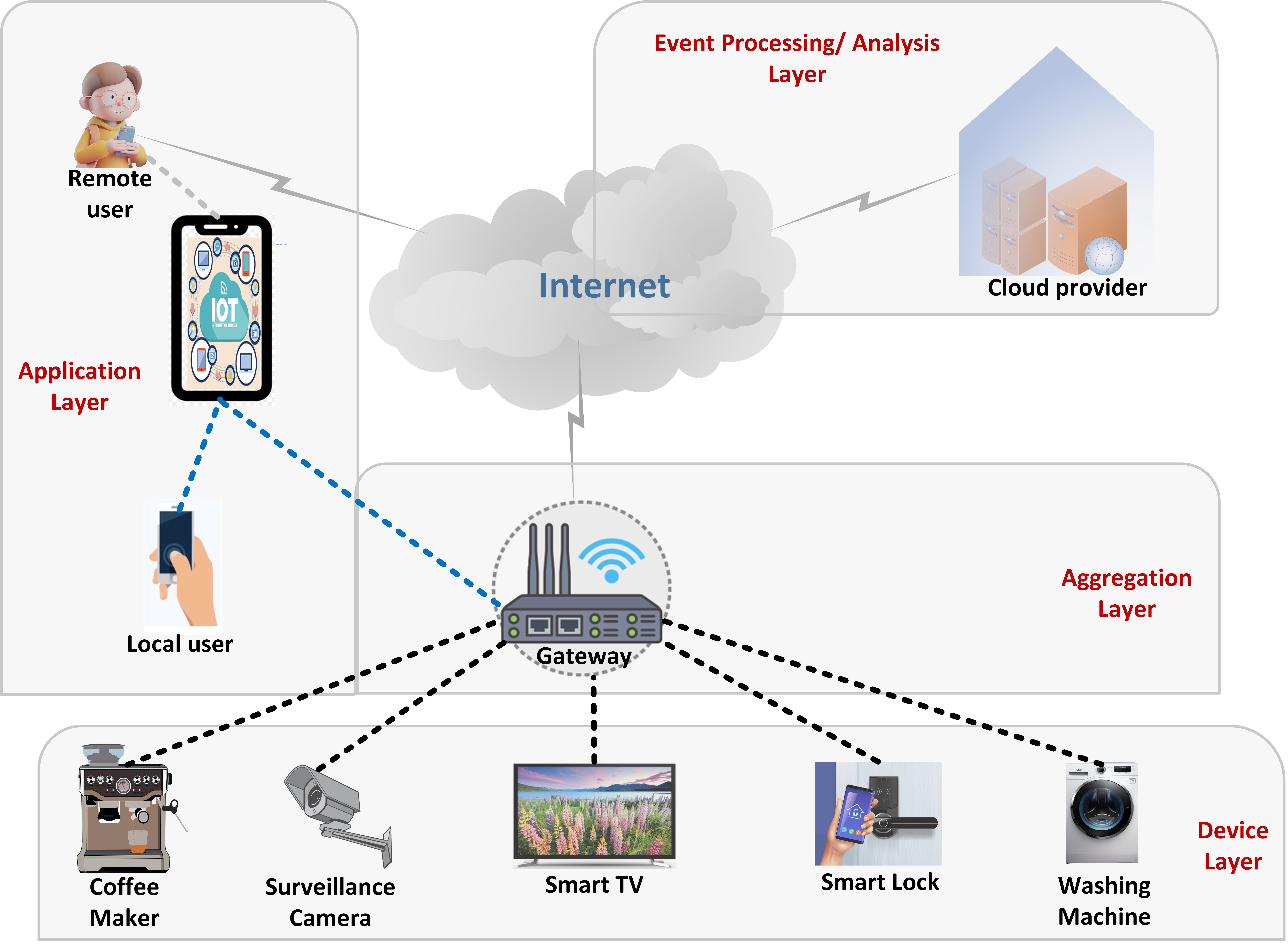}
        \caption{Layered Architecture of an IoT-Based Smart Home System Illustrating Device, Aggregation, Application, and Event Processing Layers with IoT devices, User, Gateway, and Cloud Interactions represented in each layer}
        \label{fig:SH-architecture}
    \end{figure}

In light of these research limitations, this study will focus on analyzing, assessing, and ranking privacy threats to both the user and the smart device in the Smart Home (SH), using the LINDDUN PRO Privacy Engineering (PE) framework by DistriNet Research Unit \cite{LINDDUN} as shown in \autoref{fig:LINDDUN-framework}. Leveraging this LINDDUN PRO PE framework in \autoref{fig:LINDDUN-framework}, we begin with the first stage, which is called \textit{model the system} by presenting a representation of the SH architecture, as shown in \autoref{fig:SH-architecture}. Based on this SH architecture, we develop a reference model that shows the layers, phases, and activities of the SH system. To create a more granular representation of the components (the entities, processes, data storage, and data flow) within the SH system, we present a DFD from the reference model. The DFD effectively illustrates the interaction of components within the Smart Home (SH). The second stage, called \textit{elicitation of threats}, involves the Privacy Threat Analysis (PTA) using the LINDDUN PRO threat model \cite{LINDDUNTree}. The third stage of the LINDDUN PRO PE framework called \textit{threat management} includes the Privacy Impact Assessment (PIA) process that evaluates the probability of a threat occurring with its consequence, prioritizes threats based on their impact on the SH system, and presents Privacy-Enhancing Technologies (PETs) that can mitigate these privacy threats \cite{LINDDUN}. 

The Data Flow Diagram (DFD) in this study encompasses (1) user registration and access management, addressing user privacy management, and (2) smart device integration and activity management, focusing on device privacy management. These processes represent our approach to addressing the limitations of previous research on device privacy preservation.

\textbf{Contribution:} The application of PE in the SH system is presented in this work by providing common privacy threats for SH from the literature and where privacy can be implemented, operated, and maintained in SH. To the best of our knowledge, this work is the first to apply the LINDDUN PRO PE framework to implement PE in the SH system domain (see \autoref{tab:comparison}). The following are the main contributions of our work.

\begin{itemize}

     \item This work introduces a reference model for the SH system that describes the core entities, their interactions, and the process phases integral to SH operations. Based on this reference model, a DFD is developed to visually represent the movement of data between the system components, thus facilitating a comprehensive understanding of information processing within the SH system.
     
    \item We conducted an in-depth PTA of the SH system utilizing the LINDDUN PRO threat model to systematically analyze privacy threats specific to the SH system, as identified in the literature, rather than relying solely on the general threats represented by the LINDDUN acronym. This analysis adopts a source-flow-destination methodology to map potential privacy threats to the interactions among DFD elements, thereby providing a structured and comprehensive assessment of privacy risks within the SH system.

    \item  We express PIA in our study as the privacy risk and prioritize these privacy threats based on their privacy risk using the expression of the privacy risk by NIST.

    \item We validate the impact of PET implementation in reducing the privacy risk in the SH system. Then, suggest other possible PETs that can mitigate some of these privacy threats in the SH system
\end{itemize}

\textbf{Organization}: The rest of the paper is organized as follows. Section \ref{SH-Overview} explains the primary privacy threats in Smart Home (SH), clarifies the relationship between PE, PTA, and PIA, and provides an overview of the LINDDUN PRO PE framework and LINDDUN PRO threat model.  This section also details the foundational process of PE implementation in our study, including the use of the LINDDUN PRO threat model template for identifying and documenting privacy threats, the reference model, and the associated Data Flow Diagram (DFD). Section \ref{section:related work} reviews relevant literature and presents a comparative analysis table to highlight the unique contributions of our approach. Section \ref{section: PE} explains the methodology of our proposed PE approach in SH system explaining both the reference model and DFD, and demonstrates how the LINDDUN threat model template is used to analyze, identify, and document privacy threats by mapping each threat to DFD elements in \autoref{section: Reference Model} and \autoref{section: Elicitation}. Section \ref{section:PIA} focuses on privacy threat management through the calculation of the PIA for each privacy threat and prioritizing these threats based on their impact. Section \ref{section: Validation} discusses the validation of our work by illustrating how the implementation of PET can reduce the privacy risk across the SH system. Section \ref{section:Discussion} provides a critical discussion of the PE framework, its real-world applicability, limitations, and trade-offs. Finally, Section \ref{section:conclusion} concludes the paper by summarizing the research findings and suggesting directions for future work.

\section{Overview of Privacy Concepts and Terms in Smart Home (SH) System}
\label{SH-Overview}
In this section, we discuss the privacy threats associated with Smart Home (SH) systems and the concept of Privacy Engineering (PE), particularly in relation to Privacy Impact Assessment (PIA) and Privacy Threat Analysis (PTA). Furthermore, we explain the LINDDUN PRO PE framework alongside the LINDDUN threat model with its template. Finally, we describe the reference model and the Data Flow Diagram (DFD) employed in this study.

 \subsection{Privacy Threats in Smart Home (SH)}
 Privacy threats are activities, malicious or not, that expose and misuse sensitive data by creating or exploiting existing vulnerabilities within a SH system \cite{bugeja2017analysis}. The following is the outline of privacy threats specific to SH systems and their respective operations, as gathered from relevant literature.

 \begin{itemize}
     \item T1-Identification of SH user: The ability of a threat actor to associate an individual within the IoT application, such as a SH system, with collected sensitive information that can pose a significant privacy risk. This privacy threat arises mainly from the lack of data minimization practices, which can lead to inadvertent disclosure of excessive information \cite{seliem2018towards, ziegeldorf2014privacy, ogonji2020survey, shaikh2019internet, alhalafi2019privacy, al2016overview, jain2020privacy}.

     \item T2-Identification of SH device: The ability of a privacy threat actor to identify the type and status of an IoT device through gathered sensitive information related to the device, thereby exposing the device's known vulnerabilities. For example, a smart device with Stateless Address Autoconfiguration (SLAAC) autoconfigures IPv6 has a predictable part called an EUI-64 format address generated from its MAC address. This MAC address is a piece of sensitive information about the smart device, which could reveal its identity \cite{hu2024iot}.  
     \item T3-Localization and Tracking: It is the continuous monitoring of a user or smart device's location over time and gathering information about their movements. This activity is usually done without the users' or smart devices' consent. For example, user tracking using Global Positioning System (GPS) as explained by the author in \cite{manoharan2009gps, seliem2018towards, ziegeldorf2014privacy, ogonji2020survey, shaikh2019internet, alhalafi2019privacy, al2016overview, jain2020privacy}.
     \item T4-Profiling: This privacy threat leverages the compiled and correlated data about users' activities, which has been analyzed to infer the behavior and personal interests of users. Online profiling is an example of this privacy threat in e-commerce, where organizations use it to know their customer for advertising and service matching to promote marketing \cite{seliem2018towards, ziegeldorf2014privacy, ogonji2020survey, shaikh2019internet, alhalafi2019privacy, al2016overview, jain2020privacy}.
     \item T5-Impersonation: It occurs when the identity of an authorized user is stolen and used for illegal purposes. This event poses significant risks and jeopardizes the confidentiality and integrity of the stolen information as a result of impersonation \cite{zainuddin2021study}.
     \item T6- Linkage of SH user data: This privacy threat involves revealing actual information about a user or smart device by connecting data from different sources. This connected information is usually isolated within individual sources but gets linked together during a collaboration process. For example, the inevitable bypass of privacy protection mechanisms during system collaboration. This allows unauthorized access and linking of privacy information during this collaboration  \cite{ziegeldorf2014privacy, ogonji2020survey, shaikh2019internet, al2016overview, jain2020privacy}.
    \item T7-Linkage of SH device data: Smart devices have unique communication patterns such as packet size, timing, and protocol usage, which can be used to identify these smart devices through their network traffic fingerprinting \cite{chowdhury2024communication}. Also, some characteristic behavior of smart devices based on their function (for example, surveillance camera vs smart door) can reveal their identity \cite{chowdhury2023internet} \cite{kostas2022iotdevid}. The aggregation of these unique features could increase the likelihood of identifying the smart device in the SH system.
     \item T8-Data Leakage: This privacy threat results from the unaware exposure of sensitive information about a system, user, or smart device. Due to the massive amount of data generated by IoT devices and the sensitivity of the data, data leakage could occur during data storage, transmission, and sharing if privacy protective measures are not in place \cite{zainuddin2021study, song2024apldp}.
     \item T9-Jurisdiction Risk: The outsourcing of cloud-based applications and services by service providers to multiple parties has brought about an elevated risk of disclosing sensitive information. In this scenario, the service provider often shares user information with third-party companies without the user's knowledge or consent \cite{zainuddin2021study}.
     \item T10-Life Cycle Transition: After smart devices become obsolete and are no longer in use, they often retain sensitive information from their previous systems. In the wrong hands, this information can be exploited for malicious purposes. For example, videos, audio recordings, or photos stored on unused cameras pose a privacy threat to the individuals whose data is involved \cite{ziegeldorf2014privacy, ogonji2020survey, shaikh2019internet, alhalafi2019privacy, al2016overview, jain2020privacy}.
     \item T11-Inventory Attack: Illegitimate gathering of information about the features, identity, and existence of SH entities such as user and smart device. Due to the interconnections of these entities in SH and their accessibility through the internet, they can be queried from anywhere by both authorized and unauthorized users \cite{ziegeldorf2014privacy, ogonji2020survey, shaikh2019internet, alhalafi2019privacy, al2016overview}.
    
 \end{itemize}

\subsection {The Correlation Between Privacy Engineering (PE), Privacy Impact Assessment (PIA), and Privacy Threat Analysis (PTA)}
\label{sec:documatation}
To better understand the concepts of PE, PTA, and PIA used in this study and their relationship with each other, we explain each concept as presented by privacy regulatory standards and relevant scholarly literature. 

\textbf{Privacy Engineering (PE)} is a field dedicated to developing methods and techniques that ensure privacy within information systems. Leading organizations such as the National Institute of Standards and Technology (NIST) and the International Association of Privacy Professionals (IAPP) emphasize the importance of risk reduction and the responsible use of data through frequent privacy management in a system \cite{nist} \cite{iapp}. Moreover, the concept of privacy has progressed beyond the initial idea of Privacy by Design (PbD) and has encompassed a more frequent privacy management process. PE presents an ongoing process that integrates privacy considerations at every stage of the Information and Communication Technology (ICT) system lifecycle \cite{williams2022privacy}. This process involves conducting PTA and PIA to strengthen system privacy \cite{kung2017privacy,olukoya2022assessing}. A notable example is the LINDDUN PRO PE framework, which adopts a privacy-centric approach by embedding PTA and PIA into the system’s privacy engineering implementation \cite{LINDDUN}. The LINDDUN PRO PE framework offers comprehensive guidelines for systematically analyzing, identifying, and documenting potential privacy threats within a system.

\textbf{Privacy Impact Assessment (PIA)} is crucial to evaluate the privacy implications of processes, policies, or products that handle personal information. PIA involves ongoing learning and anticipating potential privacy threats from emerging technologies \cite{wright2012privacy}. It stands as a fundamental approach within PE, aiding in risk mitigation, trust-building, liability prevention, and early detection of system vulnerabilities \cite{robles2020linddun, wright2012privacy}. Moreover, one of the pillars of achieving effective PIA is Privacy Threat Analysis (PTA), as explained by Robles-Gonzales et al. \cite{robles2020linddun}.

\textbf{Privacy Threat Analysis (PTA)} involves presenting and implementing privacy threat modeling to analyze a particular system for possible privacy threats, the likelihood of occurrence of these threats, and their area of occurrence. An example of a PTA model is the LINDDUN PRO threat model \cite{LINDDUNTree}, which we use in our study for the elicitation of privacy threats, and it is the most well-known efficient privacy threat model \cite{robles2020linddun}. LINDDUN is an abbreviation that represents Linking (L), Identifying (I), Non-repudiation (Nr),  Detecting (D), Data Disclosure (D), Unawareness (U), and Non-compliance (N). The abbreviation explains the general privacy threats in any system, and this is further explained in the release of the LINDDUN PRO privacy threat modeling tutorial \cite{LINDDUNTree, lindduntutorial}.  LINDDUN PRO threat model elicits privacy threats through element-based or interaction-based methods \cite{sion2018interaction}. The source-flow-destination feature in an interaction-based approach aids the improvement in threat analysis results. This feature makes the interaction-based approach more effective in Privacy Threat Analysis (PTA) than the element-based approach. 

To initiate the elicitation of privacy threats within a system using the LINDDUN privacy threat model, it is essential to identify and systematically document these threats through the LINDDUN PRO threat model template \cite{LINDDUNTree}. In this study, the template is employed to capture and record privacy threats pertinent to both SH users and smart devices. Presented below is a generic example of the LINDDUN PRO threat model template as applied to a SH system. This template serves as a foundational framework that can be tailored to address specific privacy threats within the SH system.

\begin{enumerate}[label=(\alph*)]
    \item \textbf{Misactors}: These are actors involved in the misuse of cases that threaten the privacy of the SH system. In our study, we present eight misactors in SH, namely;
    \begin{itemize}
        \item \textit{Unskilled insider}: They are usually curious authorized users who want to try something new on the SH system, creating vulnerabilities that privacy threats can exploit. Misconfiguration is a common example that can pose a threat to the privacy of the SH system \cite{nurse2014understanding, casey2007threat}.
        \item \textit{Skilled insider}: They are users who have privileged access to the network, but usually with bad intentions on the targeted system \cite{nurse2014understanding, casey2007threat}
         \item \textit{Skilled outsider(Adversary)}: They usually gain access by exploiting any vulnerabilities in the system using sophisticated methods or tools. They deliberately misuse the privilege they gain in the system \cite{casey2007threat, arogundade2012towards, walton2006balancing} 
         
        \item \textit{Security agents}: These are authorized legal groups that can access and extract sensitive data from a system without the user's consent for the purpose of investigation \cite{bugeja2017analysis, casey2007threat}.
        \item \textit{Government authorities}: They also perform similar acts as the security agent but usually extract data across nations without the knowledge of the data owners  \cite{meltzer2015nternet}
        \item \textit{Service providers (SP)}: They gathered data from the subscribed user for personal gain. This act can be seen when SP uses users' data for marketing and commercial purposes. For example, SP (s) usually profile and track subscribed user activities without their consent \cite{LINDDUNTree, casey2007threat}.
        \item \textit{Third-party provider}: They received shared user data from the SP during collaboration to provide services to users. These shared data usually include sensitive data without the users' consent on the usage or purpose of data storage. \cite{ulltveit2016secure,casey2007threat}

    \end{itemize}
    \item \textbf{Asset}: They are valuable assets, such as information, processes, networks, operation personnel, and systems, that need proper handling and protection against malicious activities \cite{standard2015iso}.
    \item \textbf{Consequence}: This is the level of impact or damage experienced by assets in a system if a privacy threat eventually becomes successful \cite{nurse2017security}.
    \item \textbf{DFD element}: These are threat surfaces of the DFD that can be exploited by privacy threats due to their vulnerabilities \cite{lindduntutorial}.
   
\end{enumerate}

These components constitute the LINDDUN PRO threat model analysis template. However, not all steps of the template apply to our analysis. Applying this model template offers a systematic framework for identifying and understanding a range of privacy threats and their possible vectors of occurrence. For instance, privacy threats may originate from misactors targeting specific data assets through exploitable DFD elements. Successful exploitation attempts could yield diverse consequences depending on the targeted DFD element's functional role and inherent security measures.

\subsection{Reference Model}
 A typical Smart Home (SH) architecture comprises smart devices connected through sensors to the gateway that allows interactions among these smart devices. However, a reference model is essential to represent the components that make up the SH for a comprehensive privacy threat analysis. In the context of IoT, Ziegeldorf \textit{et al.} \cite{ziegeldorf2014privacy} proposed a typical reference model to represent a detailed IoT architecture that included relevant entities and data flow. 
 
\subsection{Data Flow Diagram (DFD)}
 A DFD represents a system in graphical form. It provides a clear representation of a system, making it easily understandable by both technical and non-technical individuals. Many threat models use DFD as part of their threat modeling process. An example of these threat models is the Open Web Application Security Project (OWASP) \cite{owaspOWASPSecurity} and the STRIDE model as described by Aptori \cite{aptoriSTRIDEThreat}. A DFD has four elements, as explained in the LINDDUN tutorial presented in \cite{lindduntutorial}. These elements are Entity (E), which refers to the external elements of the system; Process (P), which is an element responsible for executing a process within the modeled system; Data flow(DF), which represents different pathways through which information is exchanged between system elements; and Data Store (DS), which serves as repositories for storing data, such as databases, within the modeled system \cite{lindduntutorial}.

\section{Related Work}
\label{section:related work}

Research on privacy has recently attracted researchers' attention, with most studies using the LINDDUN threat model in their analysis. Wuyt et al. \cite{wuyts2014empirical} stated that LINDDUN is a promising privacy threat model essential for privacy management in any system, and many studies have employed the LINDDUN threat model in different forms for privacy management. 

In their application of LINDDUN in a communication system, Hafbauer et al. \cite{hofbauer2012conducting} presented a privacy impact analysis to analyze communication records in  Voice-over-IP calls using LINDDUN. Their outcome provides a systematic pattern-based identification of privacy requirements for communication records and the ease of identifying privacy measures that meet the requirements. The assessment improves the privacy preservation of participant callers during the security analysis of communication records by encouraging the reuse of structured techniques. The structured technique used in this work is a PIA method used to analyze and obtain privacy requirements to support the lawful operation of communication security systems, and provides a systematic identification of relevant privacy threats, and presents appropriate mitigation techniques \cite{hofbauer2012conducting}. Their approach utilizes the LINDDUN threat model solely to conduct a privacy impact analysis, which is a type of Privacy Threat Analysis (PTA) that examines privacy threats within a communication system, with a primary focus on protecting user data. However, their methodology does not incorporate a Privacy Impact Assessment (PIA), nor does it prioritize the identified threats. This omission significantly restricts their ability to address the most critical privacy risks within the communication system. 

Similarly, Nweke \textit{et al.} \cite{nweke2022linddun} investigated the integration of privacy requirements and principles into the National Identity (NID) management system. The LINDDUN threat model was deployed for threat identification in some selected scenarios in their study. The authors then used prevention techniques after the analysis using LINDDUN. They also recommend suitable privacy enhancement solutions for these identified threats \cite{nweke2022linddun}. The work aims to raise awareness of privacy threats in the NID management system, ensuring the protection of PII and compliance with relevant privacy regulations. Although their work incorporates the prioritization of privacy threats following the completion of PTA, they do not provide details regarding the specific prioritization method employed. Additionally, their study is limited to user privacy and does not include a PIA. Consequently, their methodology lacks the systematic evaluation and risk management processes that a PIA provides, resulting in an incomplete approach to privacy threat mitigation.

In a social network, Identification and Authentication (IA) usually require the disclosure of sensitive personal information to trusted and untrusted service providers during security procedures to access online resources. Robles-Gonzalez et al. \cite{robles2020linddun} applied LINDDUN to perform PTA in the IA procedure in an online system and the verification process by mapping the privacy threats to the interaction between elements in DFD. Their work aims to reveal some threats to IA procedures in an online system. An IA model was developed from this mapping that shows a detailed user data repository called user data-info based on the LINDDUN framework. Their work focuses on protecting user privacy in an online system and also addresses the security concerns and privacy impact evaluation during the IA process \cite{robles2020linddun}. PTA was carried out in this study, focusing on the preservation of user data protection. Despite employing the LINDDUN framework, they employ the first two stages of the LINDDUN framework. They did not elicit, nor did they assign a priority level to the privacy threats identified and assessed in their analysis.

One of the problems associated with privacy threats is the problem of threat explosion, which is usually related to systematic threat elicitation methods, making privacy analysts, engineers, and architects face a large set of potential privacy threats. To address this issue, Wuyt \textit{et al.}  \cite{wuyts2018effective} initiated questionnaires from existing LINDDUN threat trees. The questions were filtered by focusing on domain-specific knowledge. The result shows that domain-specific knowledge can improve the efficiency and effectiveness of a systematic threat modeling approach \cite{wuyts2018effective}. Although their work encompasses both PTA and PE, it remains unclear whether their primary focus is on the preservation of user data, smart device data, or both.

Furthermore, Chah et al. \cite{chah2022privacy} presented a PTA on connected and autonomous vehicles (CAV) to manage privacy in an autonomous vehicle. The analysis reveals the vulnerability that can lead to privacy risks and requirements to maintain trust and secure communications in CAV Onboard / offboard, especially in misuse situations \cite {chah2022privacy}. This novel work analyses privacy threats in CAV using a structured PTA focusing on GDPR privacy requirements. Similarly, Raciti et al. \cite{raciti2023threat} proposed a Privacy Threat Analysis (PTA) that focuses on the specific automotive domain. They presented an approach that combines Domain-Independent Threat elicitation, Domain-Dependent Asset collection, and Domain-Dependent Threat Elicitation to provide soft privacy that focuses on the appropriate use and sharing of user data while allowing the user the right to control their data.  Both  Chah et al. \cite{chah2022privacy} and  Raciti et al. \cite{raciti2023threat} centered on autonomous vehicle privacy management. They leverage the LINDDUN threat model to perform PTA for user privacy management. They did not cover PIA or prioritize the privacy threats in their study. Moreover, in their analysis of user-centric security and privacy threats in connected vehicles, Stingelová et al. \cite{stingelova2023user} utilize a comprehensive multi-model threat modeling approach. This approach integrates STRIDE for security threats, LINDDUN for privacy threats, and Security Cards to help identify unconventional attack vectors. The study uses an element-based approach that maps both user data threats to specific architectural components and data flows within the automotive IoT domain, providing a detailed overview of potential vulnerabilities associated with modern connected vehicle systems \cite{stingelova2023user}. However, the authors do not perform a full PIA, as their analysis focuses on identifying and mapping threats without progressing to threat prioritization or the development of mitigation strategies.

In the health system, protecting users' personal health information (PHI) is essential. Cui et al. \cite{pop00001} performed a PTA using the Republic of China Personal Health Information Code (PHI-Code) to protect the PHI of users. The PTA was based on the LINDDUN threat model. The work offers a detailed and structured approach to enhancing privacy protection in public health information systems. Similarly, the user-centric privacy engineering approach was proposed by Barhamgi et al. \cite{barhamgi2018user} to manage privacy in an IoT system. Their work highlights user privacy concerns as the primary focus of their work and explains this as a barrier to the adoption of smart systems. In their case study, they used an innovative privacy threat monitoring system that included the SH and a healthcare monitoring system. The result demonstrates how users can be empowered in making privacy decisions and calculating privacy risk-benefit tradeoffs in a smart environment \cite{barhamgi2018user}. Both Cui et al. \cite{pop00001} and Barhamgi et al. \cite{barhamgi2018user} address user privacy preservation by applying the LINDDUN threat model to conduct privacy threat analysis (PTA). However, only Barhamgi et al. \cite{barhamgi2018user} extend their study to include measurement of privacy risk in the health system. While Barhamgi et al. make an effort to evaluate privacy risks, their assessment focuses primarily on the sensitivity of user data, the nature of data recipients, and the extent of data exposure, rather than directly assessing the privacy risks associated with the specific threats identified in their analysis. Additionally, they did not specify how they implement PE in their work.

Brachmann et al. \cite{brachmann2023toward} analyze privacy threats in eXtended Reality (XR) systems, focusing on localization and mapping processes conducted on-device or through edge and cloud infrastructures. They use a customized threat model to address the expanded risks from offloading sensitive user and environmental data, employing the LINDDUN privacy threat analysis methodology to systematically identify privacy risks across seven categories \cite{brachmann2023toward}. While the paper does not formally conduct a PIA, it effectively discusses privacy threats and mitigation strategies. However, it does not rank these threats by likelihood or impact, concentrating specifically on the privacy implications of data offloaded to external servers.

A review of existing literature reveals that no prior studies utilizing the LINDDUN PRO threat model have comprehensively addressed Privacy Threat Analysis (PTA) through an interaction-based approach. Instead, previous analyses have predominantly relied on element-based methods, as summarized in \autoref{tab:comparison}. To mitigate the risk of a privacy threat explosion, our study confines the PTA to the Smart Home (SH) system, covering user registration and access, third-party access, and device commissioning and activity management processes. We employ a modified LINDDUN threat model tree template specifically tailored to the SH context, as detailed in \autoref{section:template} and \autoref{table:Summarize-tab}.

A distinctive aspect of our work is the implementation of a Privacy Impact Assessment (PIA) based on both the likelihood of each privacy threat occurring and the severity of its potential consequences in the context of a SH system. This approach enables us to systematically prioritize privacy threats according to their assessed risk levels, a methodological advancement over prior studies. Moreover, some previous studies that incorporated prioritization do not give a detailed explanation of the prioritization process used in their study. However, our study is unique in explicitly detailing the threat prioritization process utilized.

Furthermore, our PE application is based on the LINDDUN PRO PE framework. It is contextualized explicitly for SH systems, distinguishing our contribution from earlier studies. As highlighted in \autoref{tab:comparison}, there remains a significant gap in the literature, with most research focusing primarily on user data privacy while overlooking the benefit of device data privacy (especially device data that can be used to identify and compromise the device if the data is exposed). To address this, our study analyzes privacy threats affecting user and device data within the SH system. This systematic approach sets our work apart as a significant advancement in the field.

\begin{table*}[ht!]
\fontsize{7.2pt}{7.2pt}\selectfont
\centering
\rowcolors{2}{gray!10}{white}
\caption{Comparison of existing studies on privacy threat management across various domains, comparing the use of the LINDDUN framework, Privacy Threat Analysis (PTA) methods, Privacy Impact Assessment (PIA), Privacy Engineering (PE), consideration of user and smart device data protection, threat prioritization, and application domains.}

\label{tab:comparison}
\begin{tabularx}{\textwidth}{lcccccccX}
\toprule
\textbf{Paper} & \textbf{LINDDUN} & \textbf{PTA Method} & \textbf{PIA} & \textbf{PE} & \textbf{User Data} & \textbf{Smart Device Data} & \textbf{Threat Prioritization} & \textbf{Domain} \\
\midrule
\cite{hofbauer2012conducting}    & \checkmark &  Element-based  &  $-$       &  $-$         & \checkmark &   $-$        &   $-$        & \scriptsize Communication System \\
\cite{nweke2022linddun}          & \checkmark & Element-based   &    $-$     & $-$          & \checkmark &    $-$       & \checkmark   & \scriptsize Identification System \\
\cite{robles2020linddun}         & \checkmark & Element-based   &   $-$      &  $-$         & \checkmark &    $-$       &   $-$        & \scriptsize Social Network \\
\cite{wuyts2018effective}        & \checkmark & Element-based   &     $-$    & \checkmark   &  $\star$   &   $-$        &     $-$      & $\star$ \\
\cite{chah2022privacy}           & \checkmark & Element-based   &    $-$     &   $-$        & \checkmark &  $-$         &    $-$       & \scriptsize Automotive System \\
\cite{raciti2023threat}          & \checkmark & Element-based   &    $-$     &  $-$         & \checkmark &     $-$      &    $-$       & \scriptsize Automotive System \\
\cite{stingelova2023user}        & \checkmark & Element-based   &   $-$      &    $-$       & \checkmark &       $-$    &   $-$        & \scriptsize Automotive System \\
\cite{pop00001}                  & \checkmark & Element-based   &   $-$      &    $-$       & \checkmark &    $-$       &   $-$        & \scriptsize Health System \\
\cite{barhamgi2018user}          & \checkmark & Element-based   & \checkmark & \checkmark   & \checkmark &   $-$        &    $-$       & \scriptsize Health System \\
\cite{brachmann2023toward}       & \checkmark & Element-based   &    $-$        &    $-$       & \checkmark &   $-$        &   $-$        & \scriptsize eXtended Reality \\
\textbf{Our Study}               & \checkmark & Interaction-based \textcircled{2} & \checkmark\textcircled{3} & \checkmark \textcircled{1}  & \checkmark & \checkmark   & \checkmark   & \scriptsize SH System \\
\bottomrule
\end{tabularx}

 \footnotesize       
        
        \textbf{\checkmark}: The study covers the topic comprehensively,  $~~$ \\
         \textcircled{1}: The study uses the LINDDUN PRO Privacy Engineering (PE) framework $~~$   \\
         \textcircled{2}: The study analyzes privacy threats associated with the domain of study, not the general LINDDUN privacy threats  $~~$ \\
          \textcircled{3}: The study calculates PIA based on the likelihood of occurrence and the consequences of the privacy threat  $~~$ \\
        \textbf{$\star$}: The study did not mention $~~$  \\
        \textbf{$-$}: The study did not cover the topic.
\end{table*}

\section{Proposed Privacy Engineering (PE) in Smart Home (SH) System}  
\label{section: PE}

This study employs the LINDDUN PRO PE framework (see \autoref{fig:LINDDUN-framework}) to systematically analyze and assess privacy threats within SH systems, thereby supporting continuous privacy management. The framework’s key stages (system modeling, threat elicitation, and threat management) are applied sequentially, as detailed in the following subsections
\subsection{Modeling of a SH System}
\label{section: Reference Model}
\begin{figure*}[ht!]
    \hspace{0.9cm}
        \centering
        \includegraphics[width=14cm,height=13cm]{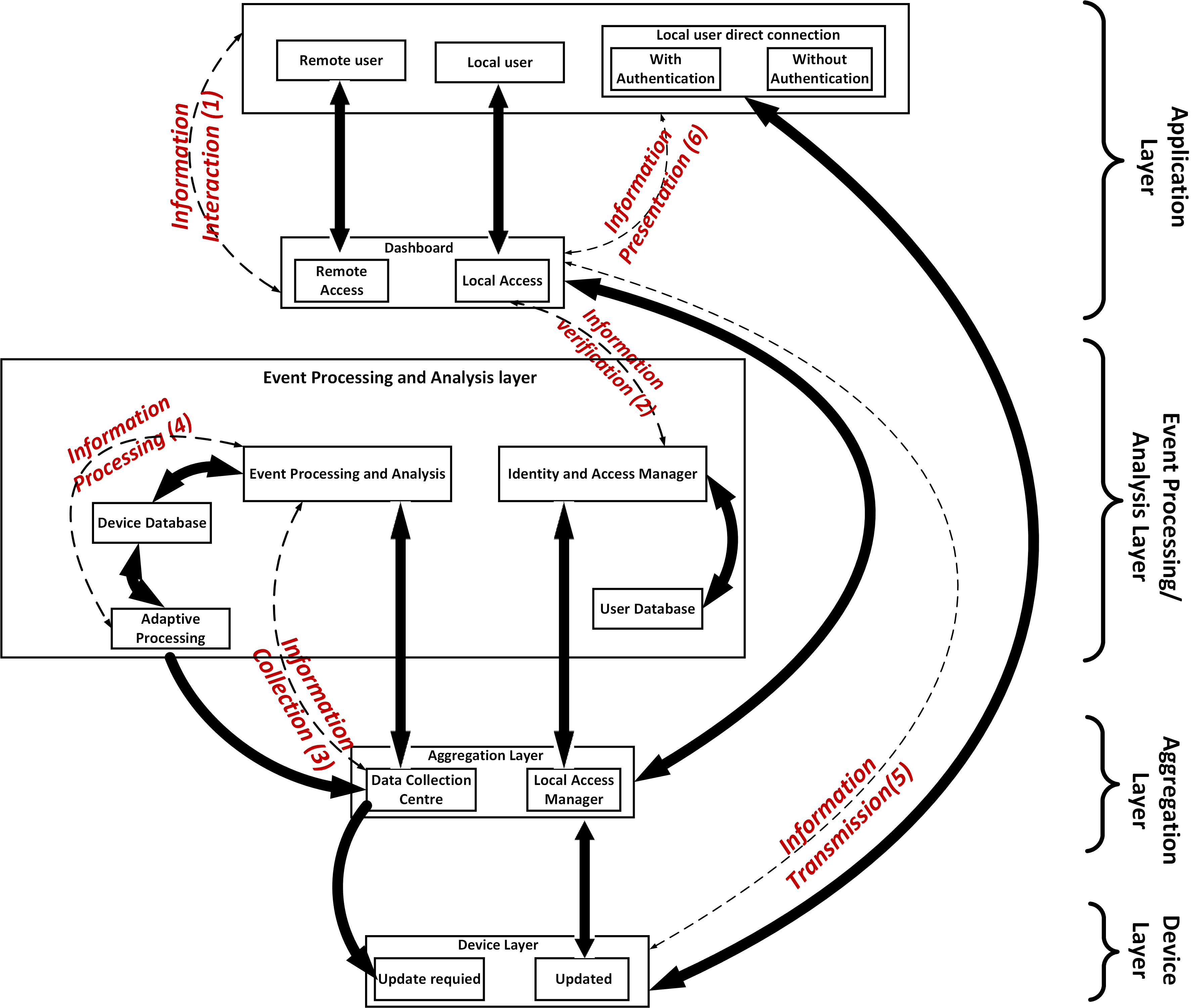}
     \caption{A reference model of a SH system illustrating the flow of information among users, devices, aggregation, and cloud-based event processing layers. The diagram details processes of interaction, verification, collection, processing, and dissemination for efficient data management.}
        \label{fig:reference-model}  
\end{figure*}

The modeling of the SH system begins with the development of a reference model that captures the essential components and interactions within a typical SH system. This reference model is further detailed using a Data Flow Diagram (DFD), which comprehensively illustrates the entities and their interactions. Building upon the reference model proposed by Ziegeldorf et al. \cite{ziegeldorf2014privacy}, this study presents an enhanced model, depicted in \autoref{fig:reference-model}, to more accurately represent a typical SH system. The layers and the communication phases in our proposed SH system reference model are explained as follows:

\begin{itemize}
    \item Layers in SH Reference Model:  The reference model proposed in this study contains four layers, as shown in \autoref{fig:reference-model}: The first layer is the \textbf{application layer} forms the top layer of the SH system, providing the user interface and the control platform of various smart devices. The second is the \textbf{event processing / analyzing layer}, which is responsible for processing the data generated by the devices, detecting events and anomalies, and making decisions based on the collected data. The third layer is the \textbf{aggregation layer}, which acts as an intermediary between the event processing/analyzing layer and the device layer, aggregating and distributing the data collected from the devices and sending some of this data to the users' dashboard in response to their request. The fourth layer is the \textbf{device layer}, which manages smart devices, including sensors, actuators, and other hardware components.
    \item Communication Phases in SH Reference Model: From our reference model in \autoref{fig:reference-model}, we identified six phases that comprise the entire activities and data flows in a typical SH system. The first phase is the \textbf{interaction phase}, where the users (remote, local, local user with direct connection) interact with the smart devices via the application programming interface (API) manager or dashboard at the system's application layer. At this phase, services such as registration and access requests that trigger user verification occur here. The second phase is the \textbf{verification phase}, in this phase, the information the user provides is verified before the user is granted access to the system through the gateway at the aggregation layer. The third phase is the \textbf{collection phase}, where the smart device data collected at the aggregation layer (gateway) is presented to the event processing manager for further analysis. The fourth phase is the \textbf{processing phase}, where the data collected is processed and analyzed to enhance the performance of smart devices, such as firmware updates. The fifth phase is the \textbf{transmission phase}, where an updated smart device now transmits the requested data to the user, presenting the data as a service to the user via the applications layer. Lastly is the \textbf{presentation phase}, where the service is provided to the user based on the instructions from the event processing/analysis layer. For a more granular explanation of these layers and communication phases, the reference model is represented in the DFD in \autoref{fig:DFD}.
\end{itemize}

\begin{figure*}[ht!]
    \hspace{0.9cm}
        \includegraphics[width=16.5cm,height=13.5cm]{DFD.jpg}
      \caption{Data Flow Diagram (DFD) illustrating the interactions and data exchanges within a SH system. The diagram maps key entities—including user devices, gateways, servers, databases, and third-party services—and details the processes of registration, authentication, commissioning, and data access regulation. The dashed lines represent data and process flows, while the red-dotted boundary line distinguishes device management from user registration processes. The legend clarifies symbols for processes, data stores, external entities, and flow types.}
        \label{fig:DFD}  
\end{figure*}

Data Flow Diagram (DFD) is introduced to illustrate the interconnections among key system components of the SH system, thereby supporting effective privacy risk management. \autoref{fig:DFD} presents the DFD for the functional architecture of a typical SH system, detailing the involved components, data flows, and processes. In this study, DFD activities are categorized into two groups: (1) device integration and activity management, and (2) user registration and access management. These categories are explained as follows.

\begin{itemize}
    \item User registration process and access management: The initial stage of the SH system involves user interaction with the application layer, where the user downloads either the Application Programming Interface (API) manager or the dashboard on their mobile device. The user then registers and logs into the SH system through the application layer. Upon receiving authentication from the authentication server and approval by the identity and access manager, the data access regulation manager determines the user's access privilege level. All user-related information is assumed to be stored in the user database for future reference. This process completes the verification phase. 

    \item Device integration and activity management: Smart devices are commissioned into the SH system in the collection phase as the first point of contact with the SH system. The smart device data are collected at the event processing and analysis layer through the Gateway located in the aggregation layer. The processed data is then sent to the smart device as an update to improve its performance. Additionally, the updated device receives users' requests through the event processing and analysis layer and responds to the request in the transmission phase. Finally, in the presentation phase, the application layer presents the user with the requested service based on the instructions of the event processing and analysis layers.
\end{itemize}

\subsection{Elicitation of Privacy Threats in Smart Home (SH) System}
\label{section: Elicitation}
Due to the lack of a suitable basis to perform a threat analysis, as mentioned by Sion \textit{et al.} \cite{sion2018interaction}, we first identify and document privacy threats activities in the SH system using the LINDDUN PRO threat model template as mentioned in \autoref{sec:documatation}. Secondly, we map each threat to interactions of elements in the DFD diagram. These two processes represent the second stage of the LINDDUN PRO PE framework. To illustrate how we document privacy threats using the LINDDUN PRO threat model template, we document the SH privacy threats, T6-linkage (SH user), in the following subsection.

 \subsubsection{Threat: T6 - Linkage (Smart Home (SH) user)}
 \label{section:template}
     This threat occurs when information from various sources can be interconnected and easily traced back to a specific SH user. This information may originate from the same or different sources, which are combined to facilitate the identification of the SH user.
\begin{itemize}
    \item{Misactor }: Skilled insider and outsider (adversary), service provider, security agency, and government authority.
    \item{Asset }: SH users' personally identifiable information (PII), such as login details, social security number, address, age, email address, personal health record, etc., can be impacted by this privacy threat if it occurs.
    \item{Consequences }: This privacy threat can aggravate threats such as T2- identification of SH user, T4-profiling, such as group discrimination as seen in racial profiling, and inference.
    \item{DFD element }: From our DFD in \autoref{fig:DFD}, SH users and any connections to the SH user that are prone to linkage privacy threats are elements that are exploitable by the misactor.
\end{itemize}
This LINDDUN PRO threat model template can be applied to other privacy threats, as summarized in \autoref{table:Summarize-tab}.  After proper identification and documentation of the privacy threats in the Smart Home(SH) system, we map these privacy threats to every interaction between the elements in the DFD diagram.

\subsubsection{Smart Home (SH) Privacy Threats Mapping to DFD Interactions}
Threat mapping is an essential process of threat analysis because it enables a detailed understanding of each data flow within a system. This study employs an interaction-based privacy threat mapping methodology, whereby interactions between elements in DFD are systematically linked to potential privacy threats. Unlike the element-based mapping approach traditionally associated with the LINDDUN threat model, our method adopts a source-flow-destination sequence to analyze interactions, offering granular insights into the likelihood of privacy threats in SH ecosystems (see  \autoref{table:map-table} and  \autoref{fig:DFD}).

We map the interactions of SH elements with specific IoT-related privacy threats from the literature that are associated with the SH system, as compared to the general threats presented by LINDDUN \cite{LINDDUNTree}. The specificity of privacy threats to the SH system in this study is a unique contribution of our work. Also, in \autoref{table:map-table}, we calculate the likelihood of the occurrence of a privacy threat by adding the number of interactions (that is, source-flow-destination) in which the privacy threat is likely to occur. This score varies among privacy threats in our study due to the focus of the privacy threat (that is, user or smart device) and the location where they could occur during the interaction. 

To illustrate our analysis, we consider the likelihood of two specific privacy threats identified in \autoref{table:map-table}. T2—identification of SH users, and T3—localization and tracking privacy threats. The T2 threat is highly likely to occur during user registration and access management, as these processes directly involve user data. In contrast, T2 is not a concern during device commissioning/activity management, since it is inherently linked to the user rather than the device.

On the other hand, the privacy threat T3—localization and tracking is likely to occur during the user registration process and access management, as well as during device commissioning/activity management. This is because both the user and the smart device can be affected by this privacy threat. For example, through wearable devices, user location can be tracked. Also, the location of stationary smart devices that are known through their sensitive data, such as IP addresses.

\autoref{table:map-table} presents the likelihood of occurrence and the corresponding privacy threat scores for each privacy threat. It is essential to note in \autoref{table:map-table} that the occurrence of T9 is relatively minimal compared to other privacy threats, whereas T6 and T11 have the highest scores. This variation is primarily attributed to the limited area where these privacy threats could occur in the SH system.

\subsection{ Privacy Threat Management in Smart Home (SH) }
\label{section:PIA}

From \autoref{table:map-table}, the scores assigned to each privacy threat likelihood of occurrence indicate their respective levels of risk within different SH elements' interactions. In particular, privacy threats T6-Linkage (SH User) and T13-Inventory attack received the highest scores, indicating a higher likelihood of occurrence. On the contrary, the T9-jurisdiction risk obtained the lowest score, suggesting a relatively lower likelihood of occurrence within SH interactions. Having calculated the likelihood of occurrence of each privacy threat, we need to calculate the privacy risk associated with these privacy threats if they occur.

In the PE support model proposed by Kung et al. \cite{kung2017privacy}, the concept of risk assessment in PE is described as quantifying privacy risks within a system. The primary objective of risk assessment is to reduce the likelihood of the occurrence of privacy threats and establish a mitigation process to counteract these threats \cite{kung2017privacy}. This process requires a comprehensive understanding of the likelihood of the occurrence of privacy threats and their impact capabilities, which is where PIA comes into play. PIA serves as a crucial component of risk assessment management in the context of PE \cite{kung2017privacy}.

Several methods of calculating the privacy risk in a system have been proposed in the literature. For example, Bloom et al. \cite{280288} presented privacy risk in their study on privacy threat modeling as a function of threats, vulnerabilities, and consequences. Furthermore, Hiller et al. \cite{hiller2017privacy} express privacy risk based on the NIST privacy framework as the likelihood of problematic data action multiplied by the impact of problematic data action. In our study, we express the privacy risk as shown in \autoref{eq:1} based on the NIST privacy framework \cite{hiller2017privacy}.

\subsubsection{Assumptions}
    The following assumptions are made in our study concerning the processes and interactions represented in the DFD in \autoref{fig:DFD}.
\begin{itemize}
    \item \textbf{A1 :} The dashboard or API manager in the DFD also performs trivial processes during communication with the IoT devices. Therefore, it is classified under the process DFD block.
    \item \textbf{A2 :}  While the cloud database (that is, the user and device database shown in \autoref{fig:reference-model} incorporates basic security measures for data storage, we assume there is a possibility of inadequate cybersecurity management and privacy policy compliance 

    \item \textbf{A3 :} Authentication of IoT users can be conducted locally through the gateway if they are connected to the local network of the IoT system
    
    \item \textbf{A4 :} The Data Access Regulation manager in the cloud is responsible for authorizing SH users, with access privileges determined based on authentication information from the gateway.
    
\end{itemize}

\begin{table*}
\fontsize{8.5pt}{8.5pt}\selectfont
\renewcommand{\arraystretch}{1.2}
\addtolength{\tabcolsep}{-6.8pt}
\caption{Comprehensive classification of privacy threats in SH system, detailing privacy threat types, potential misactors, targeted assets, consequences, and corresponding DFD elements.}
\label{table:Summarize-tab}

\begin{longtable}{
    |>{\raggedright\arraybackslash}p{2.5cm}
    |>{\raggedright\arraybackslash}p{3.0cm}
    |>{\raggedright\arraybackslash}p{2.8cm}
    |>{\raggedright\arraybackslash}p{6.3cm}
    |>{\raggedright\arraybackslash}p{2.5cm}|
}

\hline
\textbf{Privacy Threat} & \textbf{Misactor} & \textbf{Asset} &  \centering\textbf{Consequence (C)} = $I$ + $T_a$ & \textbf{DFD Elements } \\
\hline
\endfirsthead

\hline
\textbf{Privacy Threat} & \textbf{Misactor} & \textbf{Asset} & \textbf{Consequence (C)} & \textbf{DFD Elements in \autoref{fig:DFD}} \\
\hline
\endhead

\hline
\endfoot

\hline
\endlastfoot

T1 - Identification of SH element
& - Skilled insider \newline - unskilled insider
& - Smart device data \newline - gateway data.
&  i) SH element compromise ($I$ = 1) \newline ii) identified SH element could be used as an attack vector to compromise user data stores in the compromised element, resulting in T2-Identification of SH user ($T_a$ = 1). \newline  (\textbf{C=2})
& - 13- smart device(s), \newline - 12- gateway, and \newline - Any connections to 13- smart device(s) and 12- gateway
\\ \hline

T2 - Identification of SH user
&-  Skilled insiders \newline - outsiders (Adversary)
& - SH user sensitive information (PII) \newline - login details
&  i)  Compromised user data ($I$ = 1) \newline ii) The misactor instigates other privacy threats, that is T5-Impersonation and, T6-Linkage (SH user) with a stolen  ($T_a$ = 2). \newline (\textbf {C = 3})
& - SH users (1a,1b,1c) \newline - Any connections to user (1a,1b,1c)
\\ \hline

T3 - Localization and Tracking of smart devices and users
& - Service provider \newline - cloud provider \newline - security agency \newline - government.
& - SH user \newline - smart device  \newline - SH user and device location and activities
& i) To locate the culprit by the security agency, illegal monitoring, device theft, and burglary of SH ($I$ = 1) \newline ii) It can aggravate T4-profiling ($T_a$ = 1) \newline (\textbf{C=2})

&- SH users (1a,1b,1c) \newline - 12-smart device(s)
\\ \hline

T4 - Profiling
& - Service provider \newline - cloud provider \newline - skilled adversary
& - SH users
& i) Misactor carries out environment profiling and target advertisement based on this privacy threat ($I$ = 1)  \newline ii) This privacy threat can aggravate  T3-tracking privacy threats ($T_a$ = 1). \newline  (\textbf{C=2})
& - SH users (1a,1b,1c) \newline - 12-SH device(s) \newline 12-Gateway
\\ \hline

T5 - Impersonation
& - Skilled insider \newline - Unskilled Outsider (Adversary)
& - SH users' access credentials, such as login details
& i) This occurs as a result of stolen user access credential ($I$ = 1) \newline ii) This privacy threat can aggravate to  T1- identification of SH device, and T2-identification of SH users on the network ($T_a$ = 2)  \newline  (\textbf{ C=3})
& - SH users (1a,1b,1c)
\\ \hline

T6 - Linkage (SH User)
& - Skilled insider \newline - Skilled Outsider (Adversary) \newline - service provider \newline - Government Authority
& - SH users' personal information, such as social security and personal health records, etc.
& i) Revealing of user identity from different sources ($I$ = 1) \newline ii) This privacy threat can aggravate to T2-Identification users,and T4-profiling by inference, such as group discrimination. ($T_a$ = 2) \newline  (\textbf{ C=3})
&- SH users (1a,1b,1c) \newline  - Any connection to SH users
\\ \hline

T7 - Linkage (SH element's data)
& - Skilled insider \newline - Skilled Outsider (Adversary) \newline - Service provider \newline - Government Authority
& - Smart devices \newline - Gateway(s)
& i) Revealing device activity, states, and identity ($I$ = 1) \newline ii) This privacy threat can aggravate other privacy threats, T1- identification of SH device, and T2-identification of SH users' privacy threats ($T_a$ = 2). \newline
  (\textbf{ C = 3})
& - 13- smart device(s) \newline - 12-Gateway
\\ \hline

T8 - Data Leakage
& - Skilled insider \newline - Skilled Outsider (Adversary) \newline - Government Authority
& - Smart devices data \newline - Gateway data \newline - SH users' information.
& i) Leaked sensitive data of SH elements and SH users ($I$ = 1) \newline ii) This privacy threat can lead to Linking - (T6 and T7), T1- identification of the smart device, and T2 - identification of SH users privacy threats ($T_a$ = 5). \newline (\textbf{ C = 5})
& - 13- smart device(s)\newline - 12-Gateway \newline - SH users (1a,1b,1c)
\\ \hline

T9 - Jurisdiction Risk
& - Skilled insider \newline - Skilled Outsider(Adversary) \newline - Service provider.
& - Smart devices data\newline - Gateway data \newline - SH users' information.
& i) The jurisdiction of performing various processing activities by multiple cloud third parties at different locations increases privacy risk, especially when the process is not compliant with the policies governing cloud computing ($I$ = 1). \newline ii) This privacy threat is not likely to aggravate other threats as it is not malicious ($T_a$ = 0). \newline (\textbf{ C = 1})
& -  13- smart device(s) \newline  - 12-Gateway \newline  - SH users (1a,1b,1c)
\\ \hline

T10 - Life Cycle transition
& - Skilled Outsider (Adversary)
& - Smart devices data \newline - Gateway data \newline - SH users' information.
& i) Stolen SH user and SH device data ($I$ = 1) \newline ii) This privacy threat can aggravate to T2-identification of SH elements and T7-Linkage (SH Devices) privacy threats  ($T_a$ = 2).  \newline {(\textbf{ C = 3})}
& - 13- smart device(s) \newline -  12-Gateway
\\ \hline

T11 - Inventory attack
& - Skilled Outsider(Adversary)\newline - Security Agent \newline - Government Authority
& - Smart device data \newline - Gateway data \newline - SH users' information
& i) Gathered SH user and SH element data ($I$ = 1) \newline ii) This privacy threat can aggravate to  T1- identification of the smart device, T2-identification of SH users, T3- Localization and tracking, T4- Profiling  privacy threats ($T_a$ = 4) \newline (\textbf{ C = 5})
& - 13- smart device(s) \newline  - 12-Gateway
\\ \hline

\end{longtable}

\end{table*}

\begin{landscape}

\fontsize{6.5pt}{6.5pt}\selectfont
\addtolength{\tabcolsep}{-6.8pt}
\captionof{table}{ Privacy threat mapping table for a SH system, detailing the likelihood of occurrence ($L$) for various primary threats (T1–T11) across different interactions, including user registration, access management, device commissioning, and third-party access. The table categorizes interactions by source and process, indicates potential vulnerabilities, and calculates the likelihood of occurrence for each privacy threat to support risk assessment and mitigation strategies.}
\label{table:map-table}
\begin{longtable}{|l|l|l|l|l|l|l|l|l|l|l|l|l|l|}
\hline 
\multicolumn{3}{|p{303pt}|}{\centering {\bfseries Interaction}} & \multicolumn{11}{|p{277pt}|}{\centering {\bfseries Privacy Threat}}\\
\hline 
\multicolumn{14}{|p{600pt}|}{\centering {\bfseries User registration  and access management process}}\\ 
\hline 
\multicolumn{1}{|p{101pt}}{\raggedright Source} & \multicolumn{1}{|p{119.5510175pt}}{\raggedright Flow} & \multicolumn{1}{|p{119.5510175pt}}{\raggedright Destination} & \multicolumn{1}{|p{27pt}}{\raggedright T1} & \multicolumn{1}{|p{27pt}}{\raggedright T2} & \multicolumn{1}{|p{27pt}}{\raggedright T3} & \multicolumn{1}{|p{27pt}}{\raggedright T4} & \multicolumn{1}{|p{27pt}}{\raggedright T5} & \multicolumn{1}{|p{27pt}}{\raggedright T6} & \multicolumn{1}{|p{27pt}}{\raggedright T7} & \multicolumn{1}{|p{27pt}}{\raggedright T8} & \multicolumn{1}{|p{27pt}}{\raggedright T9} & \multicolumn{1}{|p{27pt}}{\raggedright T10} & \multicolumn{1}{|p{27pt}|}{\raggedright T11}\\ 
\hline 
\multicolumn{1}{|p{101pt}}{\raggedright User (1a,1b)} & \multicolumn{1}{|p{119.5510175pt}}{\raggedright Download request} & \multicolumn{1}{|p{119.5510175pt}}{\raggedright App Library (3)} & \multicolumn{1}{|p{29.1125pt}}{} & \multicolumn{1}{|p{27pt}}{\centering x} & \multicolumn{1}{|p{27pt}}{} & \multicolumn{1}{|p{27pt}}{} & \multicolumn{1}{|p{27pt}}{} & \multicolumn{1}{|p{27pt}}{} & \multicolumn{1}{|p{27pt}}{} & \multicolumn{1}{|p{27pt}}{} & \multicolumn{1}{|p{27pt}}{} & \multicolumn{1}{|p{27pt}}{} & \multicolumn{1}{|p{27pt}|}{}\\ 
\hline 
\multicolumn{1}{|p{101pt}}{\raggedright App Library (2)} & \multicolumn{1}{|p{119.5510175pt}}{\raggedright App Download} & \multicolumn{1}{|p{119.5510175pt}}{\raggedright Dashboard or API manger (3)} & \multicolumn{1}{|p{27pt}}{} & \multicolumn{1}{|p{27pt}}{} & \multicolumn{1}{|p{27pt}}{} & \multicolumn{1}{|p{27pt}}{} & \multicolumn{1}{|p{27pt}}{} & \multicolumn{1}{|p{27pt}}{\centering x} & \multicolumn{1}{|p{27pt}}{} & \multicolumn{1}{|p{27pt}}{} & \multicolumn{1}{|p{27pt}}{} & \multicolumn{1}{|p{27pt}}{} & \multicolumn{1}{|p{27pt}|}{}\\ 
\hline 
\multicolumn{1}{|p{101pt}}{\raggedright User (1a,1b)} & \multicolumn{1}{|p{101pt}}{\raggedright Registration query request} & \multicolumn{1}{|p{101pt}}{\raggedright Dashboard or API manger (3)} & \multicolumn{1}{|p{27pt}}{} & \multicolumn{1}{|p{27pt}}{} & \multicolumn{1}{|p{27pt}}{} & \multicolumn{1}{|p{27pt}}{} & \multicolumn{1}{|p{27pt}}{} & \multicolumn{1}{|p{27pt}}{} & \multicolumn{1}{|p{27pt}}{} & \multicolumn{1}{|p{27pt}}{} & \multicolumn{1}{|p{27pt}}{} & \multicolumn{1}{|p{27pt}}{} & \multicolumn{1}{|p{27pt}|}{}\\ 
\hline 
\multicolumn{1}{|p{101pt}}{\raggedright Dashboard or API manger (3)} & \multicolumn{1}{|p{101pt}}{\raggedright Registration request} & \multicolumn{1}{|p{101pt}}{\raggedright Gateway (12)} & \multicolumn{1}{|p{27pt}}{} & \multicolumn{1}{|p{27pt}}{\centering x} & \multicolumn{1}{|p{27pt}}{\centering x} & \multicolumn{1}{|p{27pt}}{} & \multicolumn{1}{|p{27pt}}{} & \multicolumn{1}{|p{27pt}}{\centering x} & \multicolumn{1}{|p{27pt}}{} & \multicolumn{1}{|p{27pt}}{\centering x} & \multicolumn{1}{|p{27pt}}{} & \multicolumn{1}{|p{27pt}}{} & \multicolumn{1}{|p{27pt}|}{\centering x}\\ 
\hline 
\multicolumn{1}{|p{101pt}}{\raggedright Gateway (12)} & \multicolumn{1}{|p{101pt}}{\raggedright Registration request} & \multicolumn{1}{|p{101pt}}{\raggedright Registration server (4)} & \multicolumn{1}{|p{27pt}}{} & \multicolumn{1}{|p{27pt}}{\centering x} & \multicolumn{1}{|p{27pt}}{} & \multicolumn{1}{|p{27pt}}{} & \multicolumn{1}{|p{27pt}}{} & \multicolumn{1}{|p{27pt}}{\centering x} & \multicolumn{1}{|p{27pt}}{} & \multicolumn{1}{|p{27pt}}{\centering x} & \multicolumn{1}{|p{27pt}}{} & \multicolumn{1}{|p{27pt}}{} & \multicolumn{1}{|p{27pt}|}{\centering x}\\ 
\hline 
\multicolumn{1}{|p{101pt}}{\raggedright Registration server (4)} & \multicolumn{1}{|p{101pt}}{\raggedright Data registration} & \multicolumn{1}{|p{101pt}}{\raggedright User Database (5a)} & \multicolumn{1}{|p{27pt}}{} & \multicolumn{1}{|p{27pt}}{} & \multicolumn{1}{|p{27pt}}{} & \multicolumn{1}{|p{27pt}}{} & \multicolumn{1}{|p{27pt}}{} & \multicolumn{1}{|p{27pt}}{\centering x} & \multicolumn{1}{|p{27pt}}{} & \multicolumn{1}{|p{27pt}}{} & \multicolumn{1}{|p{27pt}}{} & \multicolumn{1}{|p{27pt}}{\centering x} & \multicolumn{1}{|p{27pt}|}{}\\ 
\hline 
\multicolumn{1}{|p{101pt}}{\raggedright Registration server (4)} & \multicolumn{1}{|p{101pt}}{\raggedright Registration response} & \multicolumn{1}{|p{101pt}}{\raggedright Gateway (12)} & \multicolumn{1}{|p{27pt}}{} & \multicolumn{1}{|p{27pt}}{} & \multicolumn{1}{|p{27pt}}{} & \multicolumn{1}{|p{27pt}}{} & \multicolumn{1}{|p{27pt}}{} & \multicolumn{1}{|p{27pt}}{\centering x} & \multicolumn{1}{|p{27pt}}{} & \multicolumn{1}{|p{27pt}}{\centering x} & \multicolumn{1}{|p{27pt}}{} & \multicolumn{1}{|p{27pt}}{} & \multicolumn{1}{|p{27pt}|}{}\\ 
\hline 
\multicolumn{1}{|p{101pt}}{\raggedright Gateway (12)} & \multicolumn{1}{|p{101pt}}{\raggedright Registration response} & \multicolumn{1}{|p{101pt}}{\raggedright Dashboard or API manger (3)} & \multicolumn{1}{|p{27pt}}{} & \multicolumn{1}{|p{27pt}}{\centering x} & \multicolumn{1}{|p{27pt}}{} & \multicolumn{1}{|p{27pt}}{} & \multicolumn{1}{|p{27pt}}{} & \multicolumn{1}{|p{27pt}}{\centering x} & \multicolumn{1}{|p{27pt}}{} & \multicolumn{1}{|p{27pt}}{\centering x} & \multicolumn{1}{|p{27pt}}{} & \multicolumn{1}{|p{27pt}}{} & \multicolumn{1}{|p{27pt}|}{}\\ 
\hline 
\multicolumn{1}{|p{101pt}}{\raggedright Dashboard or API manger (3)} & \multicolumn{1}{|p{101pt}}{\raggedright Registration query request} & \multicolumn{1}{|p{101pt}}{\raggedright User (1a,1b)} & \multicolumn{1}{|p{27pt}}{} & \multicolumn{1}{|p{27pt}}{\centering x} & \multicolumn{1}{|p{27pt}}{} & \multicolumn{1}{|p{27pt}}{} & \multicolumn{1}{|p{27pt}}{} & \multicolumn{1}{|p{27pt}}{\centering x} & \multicolumn{1}{|p{27pt}}{} & \multicolumn{1}{|p{27pt}}{\centering x} & \multicolumn{1}{|p{27pt}}{} & \multicolumn{1}{|p{27pt}}{} & \multicolumn{1}{|p{27pt}|}{}\\ 
\hline 
\multicolumn{1}{|p{101pt}}{\raggedright User (1a,1b)} & \multicolumn{1}{|p{101pt}}{\raggedright Login query request} & \multicolumn{1}{|p{101pt}}{\raggedright Dashboard or API manger (3)} & \multicolumn{1}{|p{27pt}}{} & \multicolumn{1}{|p{27pt}}{\centering x} & \multicolumn{1}{|p{27pt}}{} & \multicolumn{1}{|p{27pt}}{\centering x} & \multicolumn{1}{|p{27pt}}{\centering x} & \multicolumn{1}{|p{27pt}}{\centering x} & \multicolumn{1}{|p{27pt}}{} & \multicolumn{1}{|p{27pt}}{\centering x} & \multicolumn{1}{|p{27pt}}{} & \multicolumn{1}{|p{27pt}}{} & \multicolumn{1}{|p{27pt}|}{}\\ 
\hline 
\multicolumn{1}{|p{101pt}}{\raggedright Dashboard or API manger (3)} & \multicolumn{1}{|p{101pt}}{\raggedright Login request} & \multicolumn{1}{|p{101pt}}{\raggedright Gateway (12)} & \multicolumn{1}{|p{27pt}}{} & \multicolumn{1}{|p{27pt}}{\centering x} & \multicolumn{1}{|p{27pt}}{} & \multicolumn{1}{|p{27pt}}{\centering x} & \multicolumn{1}{|p{27pt}}{\centering x} & \multicolumn{1}{|p{27pt}}{\centering x} & \multicolumn{1}{|p{27pt}}{} & \multicolumn{1}{|p{27pt}}{\centering x} & \multicolumn{1}{|p{27pt}}{} & \multicolumn{1}{|p{27pt}}{} & \multicolumn{1}{|p{27pt}|}{}\\ 
\hline 
\multicolumn{1}{|p{101pt}}{\raggedright Gateway (12)} & \multicolumn{1}{|p{101pt}}{\raggedright Login request} & \multicolumn{1}{|p{101pt}}{\raggedright Login server (7)} & \multicolumn{1}{|p{27pt}}{} & \multicolumn{1}{|p{27pt}}{\centering x} & \multicolumn{1}{|p{27pt}}{\centering x} & \multicolumn{1}{|p{27pt}}{} & \multicolumn{1}{|p{27pt}}{} & \multicolumn{1}{|p{27pt}}{\centering x} & \multicolumn{1}{|p{27pt}}{} & \multicolumn{1}{|p{27pt}}{\centering x} & \multicolumn{1}{|p{27pt}}{} & \multicolumn{1}{|p{27pt}}{} & \multicolumn{1}{|p{27pt}|}{\centering x}\\ 
\hline 
\multicolumn{1}{|p{101pt}}{\raggedright Login server (7)} & \multicolumn{1}{|p{101pt}}{\raggedright Authentication query request} & \multicolumn{1}{|p{101pt}}{\raggedright Authentication server (6)} & \multicolumn{1}{|p{27pt}}{} & \multicolumn{1}{|p{27pt}}{} & \multicolumn{1}{|p{27pt}}{} & \multicolumn{1}{|p{27pt}}{} & \multicolumn{1}{|p{27pt}}{} & \multicolumn{1}{|p{27pt}}{} & \multicolumn{1}{|p{27pt}}{} & \multicolumn{1}{|p{27pt}}{} & \multicolumn{1}{|p{27pt}}{} & \multicolumn{1}{|p{27pt}}{} & \multicolumn{1}{|p{27pt}|}{}\\ 
\hline 
\multicolumn{1}{|p{101pt}}{\raggedright Authentication server (6)} & \multicolumn{1}{|p{101pt}}{\raggedright Authentication query request} & \multicolumn{1}{|p{101pt}}{\raggedright Identity and access manager (9)} & \multicolumn{1}{|p{27pt}}{} & \multicolumn{1}{|p{27pt}}{} & \multicolumn{1}{|p{27pt}}{} & \multicolumn{1}{|p{27pt}}{} & \multicolumn{1}{|p{27pt}}{} & \multicolumn{1}{|p{27pt}}{} & \multicolumn{1}{|p{27pt}}{} & \multicolumn{1}{|p{27pt}}{} & \multicolumn{1}{|p{27pt}}{} & \multicolumn{1}{|p{27pt}}{} & \multicolumn{1}{|p{27pt}|}{}\\ 
\hline 
\multicolumn{1}{|p{101pt}}{\raggedright Identity and access manager (9)} & \multicolumn{1}{|p{101pt}}{\raggedright Authorization request} & \multicolumn{1}{|p{101pt}}{\raggedright Data access regulator (11)} & \multicolumn{1}{|p{27pt}}{} & \multicolumn{1}{|p{27pt}}{} & \multicolumn{1}{|p{27pt}}{} & \multicolumn{1}{|p{27pt}}{} & \multicolumn{1}{|p{27pt}}{} & \multicolumn{1}{|p{27pt}}{} & \multicolumn{1}{|p{27pt}}{} & \multicolumn{1}{|p{27pt}}{} & \multicolumn{1}{|p{27pt}}{} & \multicolumn{1}{|p{27pt}}{} & \multicolumn{1}{|p{27pt}|}{}\\ 
\hline 
\multicolumn{1}{|p{101pt}}{\raggedright Data access regulator (11)} & \multicolumn{1}{|p{101pt}}{\raggedright Data retrieval} & \multicolumn{1}{|p{101pt}}{\raggedright User Database (5a)} & \multicolumn{1}{|p{27pt}}{} & \multicolumn{1}{|p{27pt}}{} & \multicolumn{1}{|p{27pt}}{} & \multicolumn{1}{|p{27pt}}{} & \multicolumn{1}{|p{27pt}}{} & \multicolumn{1}{|p{27pt}}{} & \multicolumn{1}{|p{27pt}}{} & \multicolumn{1}{|p{27pt}}{} & \multicolumn{1}{|p{27pt}}{} & \multicolumn{1}{|p{27pt}}{} & \multicolumn{1}{|p{27pt}|}{}\\ 
\hline 
\multicolumn{1}{|p{101pt}}{\raggedright Data access regulator (11)} & \multicolumn{1}{|p{101pt}}{\raggedright Data retrieval} & \multicolumn{1}{|p{101pt}}{\raggedright Device Database (5b)} & \multicolumn{1}{|p{27pt}}{} & \multicolumn{1}{|p{27pt}}{} & \multicolumn{1}{|p{27pt}}{} & \multicolumn{1}{|p{27pt}}{} & \multicolumn{1}{|p{27pt}}{} & \multicolumn{1}{|p{27pt}}{} & \multicolumn{1}{|p{27pt}}{} & \multicolumn{1}{|p{27pt}}{} & \multicolumn{1}{|p{27pt}}{} & \multicolumn{1}{|p{27pt}}{} & \multicolumn{1}{|p{27pt}|}{}\\ 
\hline 
\multicolumn{1}{|p{101pt}}{\raggedright Data access regulator (11)} & \multicolumn{1}{|p{101pt}}{\raggedright Authorization response} & \multicolumn{1}{|p{101pt}}{\raggedright Identity and access manager (9)} & \multicolumn{1}{|p{27pt}}{} & \multicolumn{1}{|p{27pt}}{} & \multicolumn{1}{|p{27pt}}{} & \multicolumn{1}{|p{27pt}}{} & \multicolumn{1}{|p{27pt}}{} & \multicolumn{1}{|p{27pt}}{} & \multicolumn{1}{|p{27pt}}{} & \multicolumn{1}{|p{27pt}}{} & \multicolumn{1}{|p{27pt}}{} & \multicolumn{1}{|p{27pt}}{} & \multicolumn{1}{|p{27pt}|}{}\\ 
\hline 
\multicolumn{1}{|p{101pt}}{\raggedright Identity and access manager (9)} & \multicolumn{1}{|p{101pt}}{\raggedright Authorization query response } & \multicolumn{1}{|p{101pt}}{\raggedright Authentication server (6)} & \multicolumn{1}{|p{27pt}}{} & \multicolumn{1}{|p{27pt}}{} & \multicolumn{1}{|p{27pt}}{} & \multicolumn{1}{|p{27pt}}{} & \multicolumn{1}{|p{27pt}}{} & \multicolumn{1}{|p{27pt}}{} & \multicolumn{1}{|p{27pt}}{} & \multicolumn{1}{|p{27pt}}{} & \multicolumn{1}{|p{27pt}}{} & \multicolumn{1}{|p{27pt}}{} & \multicolumn{1}{|p{27pt}|}{}\\ 
\hline 
\multicolumn{1}{|p{101pt}}{\raggedright Authentication server (6)} & \multicolumn{1}{|p{101pt}}{\raggedright Authorization query response } & \multicolumn{1}{|p{101pt}}{\raggedright Login server (7)} & \multicolumn{1}{|p{27pt}}{} & \multicolumn{1}{|p{27pt}}{} & \multicolumn{1}{|p{27pt}}{} & \multicolumn{1}{|p{27pt}}{} & \multicolumn{1}{|p{27pt}}{} & \multicolumn{1}{|p{27pt}}{} & \multicolumn{1}{|p{27pt}}{} & \multicolumn{1}{|p{27pt}}{} & \multicolumn{1}{|p{27pt}}{} & \multicolumn{1}{|p{27pt}}{} & \multicolumn{1}{|p{27pt}|}{}\\ 
\hline 
\multicolumn{1}{|p{101pt}}{\raggedright Login server (7)} & \multicolumn{1}{|p{101pt}}{\raggedright Login response} & \multicolumn{1}{|p{101pt}}{\raggedright Gateway (12)} & \multicolumn{1}{|p{27pt}}{} & \multicolumn{1}{|p{27pt}}{\centering x} & \multicolumn{1}{|p{27pt}}{\centering x} & \multicolumn{1}{|p{27pt}}{} & \multicolumn{1}{|p{27pt}}{\centering x} & \multicolumn{1}{|p{27pt}}{\centering x} & \multicolumn{1}{|p{27pt}}{} & \multicolumn{1}{|p{27pt}}{\centering x} & \multicolumn{1}{|p{27pt}}{} & \multicolumn{1}{|p{27pt}}{} & \multicolumn{1}{|p{27pt}|}{\centering x}\\ 
\hline 
\multicolumn{1}{|p{101pt}}{\raggedright Gateway (12)} & \multicolumn{1}{|p{101pt}}{\raggedright Login response} & \multicolumn{1}{|p{101pt}}{\raggedright Dashboard or API manger (3)} & \multicolumn{1}{|p{27pt}}{\centering x} & \multicolumn{1}{|p{27pt}}{\centering x} & \multicolumn{1}{|p{27pt}}{\centering x} & \multicolumn{1}{|p{27pt}}{\centering x} & \multicolumn{1}{|p{27pt}}{\centering x} & \multicolumn{1}{|p{27pt}}{\centering x} & \multicolumn{1}{|p{27pt}}{} & \multicolumn{1}{|p{27pt}}{\centering x} & \multicolumn{1}{|p{27pt}}{} & \multicolumn{1}{|p{27pt}}{} & \multicolumn{1}{|p{27pt}|}{\centering x}\\ 
\hline 
\multicolumn{1}{|p{101pt}}{\raggedright Dashboard or API manger (3)} & \multicolumn{1}{|p{101pt}}{\raggedright Login query response} & \multicolumn{1}{|p{101pt}}{\raggedright User (1a,1b)} & \multicolumn{1}{|p{27pt}}{} & \multicolumn{1}{|p{27pt}}{\centering x} & \multicolumn{1}{|p{27pt}}{} & \multicolumn{1}{|p{27pt}}{} & \multicolumn{1}{|p{27pt}}{\centering x} & \multicolumn{1}{|p{27pt}}{\centering x} & \multicolumn{1}{|p{27pt}}{} & \multicolumn{1}{|p{27pt}}{\centering x} & \multicolumn{1}{|p{27pt}}{} & \multicolumn{1}{|p{27pt}}{} & \multicolumn{1}{|p{27pt}|}{}\\ 
\hline 
\multicolumn{14}{|p{600pt}|}{\centering {\bfseries Third-party access process}}\\ 
\hline 
\multicolumn{1}{|p{101pt}}{\raggedright Source} & \multicolumn{1}{|p{101pt}}{\raggedright Flow} & \multicolumn{1}{|p{101pt}}{\raggedright Destination} & \multicolumn{1}{|p{27pt}}{\raggedright T1} & \multicolumn{1}{|p{27pt}}{\raggedright T2} & \multicolumn{1}{|p{27pt}}{\raggedright T3} & \multicolumn{1}{|p{27pt}}{\raggedright T4} & \multicolumn{1}{|p{27pt}}{\raggedright T5} & \multicolumn{1}{|p{27pt}}{\raggedright T6} & \multicolumn{1}{|p{27pt}}{\raggedright T7} & \multicolumn{1}{|p{27pt}}{\raggedright T8} & \multicolumn{1}{|p{27pt}}{\raggedright T9} & \multicolumn{1}{|p{27pt}}{\raggedright T10} & \multicolumn{1}{|p{27pt}|}{\raggedright T11}\\ 
\hline 
\multicolumn{1}{|p{101pt}}{\raggedright Third-party (17)} & \multicolumn{1}{|p{101pt}}{\raggedright Access request} & \multicolumn{1}{|p{101pt}}{\raggedright Data access regulator (11)} & \multicolumn{1}{|p{27pt}}{} & \multicolumn{1}{|p{27pt}}{} & \multicolumn{1}{|p{27pt}}{} & \multicolumn{1}{|p{27pt}}{\centering x} & \multicolumn{1}{|p{27pt}}{\centering x} & \multicolumn{1}{|p{27pt}}{} & \multicolumn{1}{|p{27pt}}{} & \multicolumn{1}{|p{27pt}}{} & \multicolumn{1}{|p{27pt}}{\centering x} & \multicolumn{1}{|p{27pt}}{} & \multicolumn{1}{|p{27pt}|}{\centering x}\\ 
\hline 
\multicolumn{14}{|p{600pt}|}{\centering {\bfseries Device commissioning/activity management process}}\\ 
\hline 
\multicolumn{1}{|p{101pt}}{\raggedright Source} & \multicolumn{1}{|p{101pt}}{\raggedright Flow} & \multicolumn{1}{|p{101pt}}{\raggedright Destination} & \multicolumn{1}{|p{27pt}}{\raggedright T1} & \multicolumn{1}{|p{27pt}}{\raggedright T2} & \multicolumn{1}{|p{27pt}}{\raggedright T3} & \multicolumn{1}{|p{27pt}}{\raggedright T4} & \multicolumn{1}{|p{27pt}}{\raggedright T5} & \multicolumn{1}{|p{27pt}}{\raggedright T6} & \multicolumn{1}{|p{27pt}}{\raggedright T7} & \multicolumn{1}{|p{27pt}}{\raggedright T8} & \multicolumn{1}{|p{27pt}}{\raggedright T9} & \multicolumn{1}{|p{27pt}}{\raggedright T10} & \multicolumn{1}{|p{27pt}|}{\raggedright T11}\\ 
\hline 
\multicolumn{1}{|p{101pt}}{\raggedright smart device (13)} & \multicolumn{1}{|p{101pt}}{\raggedright commissioning request} & \multicolumn{1}{|p{101pt}}{\raggedright Gateway (12)} & \multicolumn{1}{|p{27pt}}{\centering x} & \multicolumn{1}{|p{27pt}}{} & \multicolumn{1}{|p{27pt}}{} & \multicolumn{1}{|p{27pt}}{} & \multicolumn{1}{|p{27pt}}{} & \multicolumn{1}{|p{27pt}}{} & \multicolumn{1}{|p{27pt}}{\centering x} & \multicolumn{1}{|p{27pt}}{} & \multicolumn{1}{|p{27pt}}{} & \multicolumn{1}{|p{27pt}}{} & \multicolumn{1}{|p{27pt}|}{\centering x}\\ 
\hline 
\multicolumn{1}{|p{101pt}}{\raggedright Gateway (12)} & \multicolumn{1}{|p{101pt}}{\raggedright commissioning request} & \multicolumn{1}{|p{101pt}}{\raggedright Event processing and analysis (10)} & \multicolumn{1}{|p{27pt}}{\centering x} & \multicolumn{1}{|p{27pt}}{} & \multicolumn{1}{|p{27pt}}{} & \multicolumn{1}{|p{27pt}}{} & \multicolumn{1}{|p{27pt}}{} & \multicolumn{1}{|p{27pt}}{} & \multicolumn{1}{|p{27pt}}{\centering x} & \multicolumn{1}{|p{27pt}}{} & \multicolumn{1}{|p{27pt}}{} & \multicolumn{1}{|p{27pt}}{} & \multicolumn{1}{|p{27pt}|}{\centering x}\\ 
\hline 
\multicolumn{1}{|p{101pt}}{\raggedright Event processing and analysis (10)} & \multicolumn{1}{|p{101pt}}{\raggedright Data upload} & \multicolumn{1}{|p{101pt}}{\raggedright Device Database (5b)} & \multicolumn{1}{|p{27pt}}{} & \multicolumn{1}{|p{27pt}}{} & \multicolumn{1}{|p{27pt}}{} & \multicolumn{1}{|p{27pt}}{} & \multicolumn{1}{|p{27pt}}{} & \multicolumn{1}{|p{27pt}}{} & \multicolumn{1}{|p{27pt}}{} & \multicolumn{1}{|p{27pt}}{} & \multicolumn{1}{|p{27pt}}{} & \multicolumn{1}{|p{27pt}}{\centering x} & \multicolumn{1}{|p{27pt}|}{\centering x}\\ 
\hline 
\multicolumn{1}{|p{101pt}}{\raggedright smart device (13)} & \multicolumn{1}{|p{101pt}}{\raggedright service request} & \multicolumn{1}{|p{101pt}}{\raggedright Gateway (12)} & \multicolumn{1}{|p{27pt}}{\centering x} & \multicolumn{1}{|p{27pt}}{} & \multicolumn{1}{|p{27pt}}{\centering x} & \multicolumn{1}{|p{27pt}}{} & \multicolumn{1}{|p{27pt}}{} & \multicolumn{1}{|p{27pt}}{} & \multicolumn{1}{|p{27pt}}{\centering x} & \multicolumn{1}{|p{27pt}}{} & \multicolumn{1}{|p{27pt}}{} & \multicolumn{1}{|p{27pt}}{} & \multicolumn{1}{|p{27pt}|}{\centering x}\\ 
\hline 
\multicolumn{1}{|p{101pt}}{\raggedright Gateway (12)} & \multicolumn{1}{|p{101pt}}{\raggedright service request} & \multicolumn{1}{|p{101pt}}{\raggedright Event processing and analysis (10)} & \multicolumn{1}{|p{27pt}}{\centering x} & \multicolumn{1}{|p{27pt}}{} & \multicolumn{1}{|p{27pt}}{\centering x} & \multicolumn{1}{|p{27pt}}{\centering x} & \multicolumn{1}{|p{27pt}}{} & \multicolumn{1}{|p{27pt}}{} & \multicolumn{1}{|p{27pt}}{\centering x} & \multicolumn{1}{|p{27pt}}{} & \multicolumn{1}{|p{27pt}}{} & \multicolumn{1}{|p{27pt}}{} & \multicolumn{1}{|p{27pt}|}{\centering x}\\ 
\hline 
\multicolumn{1}{|p{101pt}}{\raggedright Event processing and analysis (10)} & \multicolumn{1}{|p{101pt}}{\raggedright service request} & \multicolumn{1}{|p{101pt}}{\raggedright Device Database (5b)} & \multicolumn{1}{|p{27pt}}{} & \multicolumn{1}{|p{27pt}}{} & \multicolumn{1}{|p{27pt}}{} & \multicolumn{1}{|p{27pt}}{} & \multicolumn{1}{|p{27pt}}{} & \multicolumn{1}{|p{27pt}}{} & \multicolumn{1}{|p{27pt}}{} & \multicolumn{1}{|p{27pt}}{} & \multicolumn{1}{|p{27pt}}{} & \multicolumn{1}{|p{27pt}}{} & \multicolumn{1}{|p{27pt}|}{}\\ 
\hline 
\multicolumn{1}{|p{101pt}}{\raggedright Device Database (5b)} & \multicolumn{1}{|p{101pt}}{\raggedright Data control request} & \multicolumn{1}{|p{101pt}}{\raggedright Data access regulator (11)} & \multicolumn{1}{|p{27pt}}{} & \multicolumn{1}{|p{27pt}}{} & \multicolumn{1}{|p{27pt}}{} & \multicolumn{1}{|p{27pt}}{} & \multicolumn{1}{|p{27pt}}{} & \multicolumn{1}{|p{27pt}}{} & \multicolumn{1}{|p{27pt}}{} & \multicolumn{1}{|p{27pt}}{} & \multicolumn{1}{|p{27pt}}{} & \multicolumn{1}{|p{27pt}}{} & \multicolumn{1}{|p{27pt}|}{}\\ 
\hline 
\multicolumn{1}{|p{101pt}}{\raggedright Data access regulator (11)} & \multicolumn{1}{|p{101pt}}{\raggedright Data control response} & \multicolumn{1}{|p{101pt}}{\raggedright Device Database (5b)} & \multicolumn{1}{|p{27pt}}{} & \multicolumn{1}{|p{27pt}}{} & \multicolumn{1}{|p{27pt}}{} & \multicolumn{1}{|p{27pt}}{} & \multicolumn{1}{|p{27pt}}{} & \multicolumn{1}{|p{27pt}}{} & \multicolumn{1}{|p{27pt}}{} & \multicolumn{1}{|p{27pt}}{} & \multicolumn{1}{|p{27pt}}{} & \multicolumn{1}{|p{27pt}}{} & \multicolumn{1}{|p{27pt}|}{}\\ 
\hline 
\multicolumn{1}{|p{101pt}}{\raggedright Device Database (5b)} & \multicolumn{1}{|p{101pt}}{\raggedright service response} & \multicolumn{1}{|p{101pt}}{\raggedright Event processing and analysis (10)} & \multicolumn{1}{|p{27pt}}{} & \multicolumn{1}{|p{27pt}}{} & \multicolumn{1}{|p{27pt}}{} & \multicolumn{1}{|p{27pt}}{} & \multicolumn{1}{|p{27pt}}{} & \multicolumn{1}{|p{27pt}}{} & \multicolumn{1}{|p{27pt}}{} & \multicolumn{1}{|p{27pt}}{} & \multicolumn{1}{|p{27pt}}{} & \multicolumn{1}{|p{27pt}}{} & \multicolumn{1}{|p{27pt}|}{}\\ 
\hline 
\multicolumn{1}{|p{101pt}}{\raggedright Event processing and analysis (10)} & \multicolumn{1}{|p{101pt}}{\raggedright service response} & \multicolumn{1}{|p{101pt}}{\raggedright Gateway (12)} & \multicolumn{1}{|p{27pt}}{\centering x} & \multicolumn{1}{|p{27pt}}{} & \multicolumn{1}{|p{27pt}}{\centering x} & \multicolumn{1}{|p{27pt}}{\centering x} & \multicolumn{1}{|p{27pt}}{} & \multicolumn{1}{|p{27pt}}{} & \multicolumn{1}{|p{27pt}}{\centering x} & \multicolumn{1}{|p{27pt}}{} & \multicolumn{1}{|p{27pt}}{} & \multicolumn{1}{|p{27pt}}{} & \multicolumn{1}{|p{27pt}|}{\centering x}\\ 
\hline 
\multicolumn{1}{|p{101pt}}{\raggedright Gateway (12)} & \multicolumn{1}{|p{101pt}}{\raggedright service response} & \multicolumn{1}{|p{101pt}}{\raggedright Smart device (13)} & \multicolumn{1}{|p{27pt}}{\centering x} & \multicolumn{1}{|p{27pt}}{} & \multicolumn{1}{|p{27pt}}{\centering x} & \multicolumn{1}{|p{27pt}}{} & \multicolumn{1}{|p{27pt}}{} & \multicolumn{1}{|p{27pt}}{} & \multicolumn{1}{|p{27pt}}{\centering x} & \multicolumn{1}{|p{27pt}}{} & \multicolumn{1}{|p{27pt}}{} & \multicolumn{1}{|p{27pt}}{} & \multicolumn{1}{|p{27pt}|}{\centering x}\\ 
\hline
\multicolumn{3}{|p{303pt}}{\centering {\bfseries Total number of possible privacy threat occurrence in all interactions (Tn)}}& \multicolumn{1}{|p{27pt}}{\centering 7} & \multicolumn{1}{|p{27pt}}{\centering 11} & \multicolumn{1}{|p{27pt}}{\centering 8} & \multicolumn{1}{|p{27pt}}{\centering 6} & \multicolumn{1}{|p{27pt}}{\centering 6} & \multicolumn{1}{|p{27pt}}{\centering 13} & \multicolumn{1}{|p{27pt}}{\centering 6} & \multicolumn{1}{|p{27pt}}{\centering 11} & \multicolumn{1}{|p{27pt}}{\centering 1} & \multicolumn{1}{|p{27pt}}{\centering 2} & \multicolumn{1}{|p{27pt}|}{\centering 13}\\ 

\hline
\multicolumn{3}{|p{303pt}}{\centering {\bfseries Likelihood of Occurrence (L)}}& \multicolumn{1}{|p{27pt}}{\centering 0.20000} & \multicolumn{1}{|p{27pt}}{\centering 0.31429} & \multicolumn{1}{|p{27pt}}{\centering 0.22857} & \multicolumn{1}{|p{27pt}}{\centering 0.17143} & \multicolumn{1}{|p{27pt}}{\centering 0.17143} & \multicolumn{1}{|p{27pt}}{\centering 0.37143} & \multicolumn{1}{|p{27pt}}{\centering 0.17143} & \multicolumn{1}{|p{27pt}}{\centering 0.31429} & \multicolumn{1}{|p{27pt}}{\centering 0.02857} & \multicolumn{1}{|p{27pt}}{\centering 0.05714} & \multicolumn{1}{|p{27pt}|}{\centering 0.37143}\\ 
\hline

\end{longtable}

\end{landscape}

\subsubsection{Quantitative Assessment and Prioritization of Privacy Threats in Smart Home (SH) Systems}

The quantitative approach used in this study calculates the privacy risk (PIA) of each privacy threat based on its likelihood of occurrence and the consequences that would result if it were to occur. Firstly, we calculate the consequence and then the likelihood of occurrence. Secondly, we calculate the privacy risk (PIA) as the product of the likelihood of occurrence and the consequence of each privacy threat if it were to occur in the SH system. Finally, we prioritize each privacy threat based on its privacy risk value, using the impact level as outlined in \autoref{tab:impact-table}. The assessment process is as follows.

\begin{itemize}

\item \textbf{Consequence}: To assess the consequences of privacy threats in SH systems, each privacy threat is assigned an initial baseline consequence ($I$) value of 1. We also account for the ability of the privacy threat to aggravate additional privacy threats, denoted as the value of the aggravated privacy threat ($T_a$). The value of $T_a$ corresponds to the number of other privacy threats that can be triggered or intensified by the privacy threat in question. The total consequence value ($C$) of each privacy threat are expressed in  \autoref{table:Summarize-tab} and calculated using \autoref{eq:2} as follows:

\begin{equation}
\label{eq:2}
    C = I + T_a
\end{equation}
where
\begin{itemize}
    \item $I$ is the initial consequence value, which is the baseline consequence for each privacy threat, set to 1 for all privacy threats.
    \item $T_a$ is the number of aggravated privacy threats caused by the original privacy threat.
\end{itemize}

These calculated consequence values are reported in the consequences ($C$) column of \autoref{table:Summarize-tab}.

\item \textbf{Likelihood of Occurrence (L)}: It is defined as the ratio of the total number of occurrences of a given privacy threat ($T_n$) as shown in \autoref{table:map-table} to the total number of system interactions ($T_i$), as expressed in \autoref{eq:3} below:
\begin{equation}
\label{eq:3}
    L = \frac{T_n}{T_i}
\end{equation}
In this study, the total number of interactions ($T_i$) is 35. The likelihood values for each privacy threat are calculated using \autoref{eq:3} and are detailed in \autoref{table:map-table}.

\item \textbf{Privacy Impact Assessment (PIA)}

Privacy Impact Assessment (PIA), also known as privacy risk, is determined by multiplying the likelihood of a privacy threat's occurrence ($L$) by its consequence ($C$):
\begin{equation}
\label{eq:1}
    \text{Privacy risk (PIA)} = L \times C
\end{equation}

 While the calculated privacy risk values are obtained using \autoref{eq:1}. The summary of the likelihood of occurrence ($L$), consequences ($C$), the privacy risk (PIA) of these privacy threats, and the prioritization of these privacy threats based on their risk level is presented in \autoref{tab:PIA_Before_PET}. \textbf{Prioritization} allows ranking each privacy threat in the SH systems by quantitatively evaluating their potential impact based on calculated privacy risk value.

\textbf{Illustration}: We illustrate our PE approach by calculating the PIA for  T11-inventory attack using  \autoref{eq:2}, \autoref{eq:3}, and \autoref{eq:1} to determine the privacy risk (PIA) of T11-inventory attack within SH systems. The overall consequence of the  T11-inventory attack is quantified as 5, as indicated in \autoref{table:Summarize-tab}. This value is derived using \autoref{eq:2}, which incorporates both the initial consequence $I$ of the attack (assigned a value of 1) and the number of aggravated privacy threats $T_a$ associated with it (four in this instance), resulting in a cumulative consequence score of 5:
\begin{equation}
    C = I + T_a = 1 + 4 = 5
    \label{eq:consequence}
\end{equation}
where $I$ = 1 is the initial consequence and $T_a$ = 4 is the number of aggravated privacy threats for T11-inventory attack .

The likelihood of occurrence for the  T11-inventory attack is calculated as shown in \autoref{eq:likelihood}. For T11, the number of observed instances ($T_n$) is 13, while the total number of privacy threats ($T_i$) is 35, as shown in \autoref{table:map-table}. The likelihood is thus determined by dividing $T_n$ by $T_i$, yielding a value of $0.37143$:
\begin{equation}
    L = \frac{T_n}{T_i} = \frac{13}{35} \approx 0.37143
    \label{eq:likelihood}
\end{equation}

Subsequently, the PIA for the T11-inventory attack is calculated using \autoref{eq:pia}, which multiplies the likelihood ($L = 0.37143$) by the consequence ($C = 5$). This results in a PIA value of approximately $1.85$, as detailed in \autoref{tab:PIA_Before_PET}:
\begin{equation}
    \text{PIA} = L \times C = 0.37143 \times 5 \approx 1.85
    \label{eq:pia}
\end{equation}

This analytical approach is applied to all privacy threats considered in this study, and the corresponding privacy risk (PIA) values are comprehensively reported in \autoref{tab:PIA_Before_PET}. This structured approach ensures that both the probability of privacy threat occurrence and the severity of its consequences are quantitatively integrated, thereby supporting a rigorous and transparent privacy risk assessment process.

\item {\textbf{Prioritization of Privacy Threats}}

 \autoref{tab:impact-table} is the standard risk value classification used in our study, derived from the ranking of the privacy risk based on FIPS 199 potential impact as presented in the NIST publication 800-60 Volume 1 \cite{nist1}. However, our study set the maximum impact level category at 2. The maximum impact level and the range assigned to each level are distinctive to our work. The maximum impact level and range may not be the same in another SH system, as the range and maximum impact level can be flexibly categorized. \autoref{tab:PIA_Before_PET} expresses the ranking of privacy threats according to the possible impact that the privacy threat can have on a SH system.
T6- Linkage (SH User), T8- Data Leakage, and T11- inventory attack has the highest ranking, with T2 and T4 with moderate ranking, and the lowest ranked are T1, T3, T5, T7, T9, T10. 

\end{itemize}

\begin{table}[H]
    \centering
    \renewcommand{\arraystretch}{1.2}
    \fontsize{9pt}{11pt}\selectfont
    \caption{Impact levels and their associated risk potential.}
    \label{tab:impact-table}
    \begin{tabular}{|c|c|}
        \hline
        \textbf{Category} & \textbf{Risk Potential} \\ \hline
        1.00 -- 2.00      & \cellcolor{red!100}\textcolor{white}{High}      \\ \hline
        0.50 -- 0.99      & \cellcolor{yellow!100!black}Moderate            \\ \hline
        0.10 -- 0.49      & \cellcolor{blue!50}Low                         \\ \hline
    \end{tabular}
    
\end{table}

\begin{table*}[ht!]

    \centering
    \renewcommand{\arraystretch}{1.25}
    \fontsize{9pt}{11pt}\selectfont
    \caption{The prioritization of privacy threats in the SH system based on the PIA value calculated from Likelihood of occurrence ($L$) and consequence ($C$) if the privacy threat occurs}
     \label{tab:PIA_Before_PET}
    \begin{tabular}{|c|c|c|c|c|c|c|c|}
        \hline
        \textbf{Threat} & \textbf{I} & \textbf{Ta} & \textbf{C = I + Ta} & \textbf{Tn} & \textbf{L = Tn/Ti} & \textbf{PIA = L $\times$ C} & \textbf{Prioritization} \\ \hline
        T1  & 1 & 1 & 2 & 7  & 0.20000 & 0.4000 $\approx$ 0.40 & \cellcolor{blue!50}Low \\ \hline
        T2  & 1 & 2 & 3 & 11 & 0.31429 & 0.9429 $\approx$ 0.94 & \cellcolor{yellow!100!black}Moderate \\ \hline
        T3  & 1 & 1 & 2 & 8  & 0.22857 & 0.4571 $\approx$ 0.46 & \cellcolor{blue!50}Low \\ \hline
        T4  & 1 & 1 & 2 & 6  & 0.17143 & 0.3429 $\approx$ 0.34 & \cellcolor{blue!50}Low \\ \hline
        T5  & 1 & 2 & 3 & 6  & 0.17143 & 0.5143 $\approx$ 0.51 & \cellcolor{yellow!100!black}Moderate \\ \hline
        T6  & 1 & 2 & 3 & 13 & 0.37143 & 1.1143 $\approx$ 1.11 & \cellcolor{red!100}\textcolor{white}{High} \\ \hline
        T7  & 1 & 2 & 3 & 6  & 0.17143 & 0.5143 $\approx$ 0.51 & \cellcolor{yellow!100!black}Moderate \\ \hline
        T8  & 1 & 4 & 5 & 11 & 0.31429 & 1.5714 $\approx$ 1.57 & \cellcolor{red!100}\textcolor{white}{High} \\ \hline
        T9  & 1 & 0 & 1 & 1  & 0.02857 & 0.0286 $\approx$ 0.03 & \cellcolor{blue!50}Low \\ \hline
        T10 & 1 & 2 & 3 & 2  & 0.05714 & 0.1714 $\approx$ 0.17 & \cellcolor{blue!50}Low \\ \hline
        T11 & 1 & 4 & 5 & 13 & 0.37143 & 1.8571 $\approx$ 1.86 & \cellcolor{red!100}\textcolor{white}{High} \\ \hline
    \end{tabular}
\end{table*}

\subsection {Discussion of the privacy threat prioritization result}
From the privacy risk (PIA) in \autoref{tab:PIA_Before_PET}, the potential impact and the prioritization of each privacy threat are explained as follows. 

\begin{itemize}
    \item T6- Linkage (Smart Home (SH) user), T8- Data Leakage, and T11-Inventory attack primarily target the storage systems where sensitive information is stored \cite{ziegeldorf2014privacy}, which is part of the infrastructure in the SH system. The high ranking of these privacy threats, as depicted in \autoref{tab:PIA_Before_PET}, is attributed to the significant and irreparable impact they can have on the SH system. The recent increase in this privacy threat could result from technological advances \cite{nair2021privacy}, which have enabled more sophisticated and accessible hacking techniques compared to previous years \cite{hinde2014privacy, jang2014survey}. Emerging technologies such as Artificial Intelligence (AI) have lowered the entry barrier for individuals with a moderate level of knowledge to exploit these innovations for malicious activities \cite{alt2019emerging}\cite{dhirani2023ethical}. These factors contribute to the increased prevalence and severity of T6-Linkage (SH user), T8-Data Leakage, and T11-inventory attack in SH environments, leading to the source of most ransomware attacks experienced today. \\
    \item The ranking of T2 - Identification of SH user, T5 - Impersonation, and T7 - Linkage (smart devices' data)  privacy threats is moderately high, as they are more associated with SH users except T7- Linkage (smart device data), which is indirectly related to SH users as most smart devices store SH user information at some point in the SH system. These three privacy threats could be combined to attack the SH system successfully. When the attributes or data of the SH users can be linked, it could reveal the true identities of the users \cite{vatsalan2013taxonomy}. The identified user credential could be used to have legitimate access to the smart device, thereby giving the intruder further access to the core of the SH system through linking. This could allow the intruder to connect to other smart devices from the initially compromised smart device. The interconnection between these privacy threats further amplifies their severity, as exploiting them can result in successful privacy breaches in SH systems. \\

    \item Privacy threats T1 - Identification of SH Element and T10 - Life Cycle transition are closely linked to vulnerabilities found in smart devices such as lack of physical security, improper encryption, and unmanaged open ports \cite{ogonji2020survey} \cite{neshenko2019demystifying}. 
   The implementation of adequate privacy and security measures could reduce the likelihood that these privacy threats will occur and minimize their potential impact on the overall SH system. The low impact ranking suggests that security measures for smart devices effectively reduce the vulnerabilities associated with T1 and T10. These measures may include access and authentication controls, security protocols, and intrusion detection, among others, as stated by Ogonji et al. \cite{ogonji2020survey}. However, access to user data stored in any of the compromised SH elements could reveal the user's identity. Moreover, it remains essential to continuously assess and address vulnerabilities in smart devices, as the privacy threat landscape is constantly evolving. Ongoing efforts to enhance security measures and stay informed about emerging privacy threats are essential for maintaining the ongoing protection of SH systems, including the privacy of users and devices. In contrast, T3, T4, and T9 have been ranked low primarily because the intent of the misactor associated with these privacy threats is typically non-malicious. 
    \end{itemize}
The Privacy Engineering (PE) will continue to identify and analyze privacy threats that can directly and indirectly impact the SH system. This proactive approach to privacy ensures users' confidence in using the SH system.

\section{Evaluating the Effectiveness of PETs in Reducing Privacy Risks during User Access Management and Smart Device Commissioning/Activity Management process in SH system: A Case Study}

\label{section: Validation}

To validate the proposed Privacy Engineering (PE), we consider two distinct scenarios. In the first scenario, a PET is implemented in the user and device databases to preserve sensitive data at rest. In the second scenario, we assume the application of end-to-end encryption, utilizing an appropriate PET, to ensure the protection of user and device-sensitive data during transmission.

\subsection{Implementing a PET in User Database and End-to-end Encryption during User Access Management and Device Commissioning/Activity Management Process}
 
\begin{figure*}[ht!]
        \centering
        \includegraphics[width=17cm,height=8.5cm]{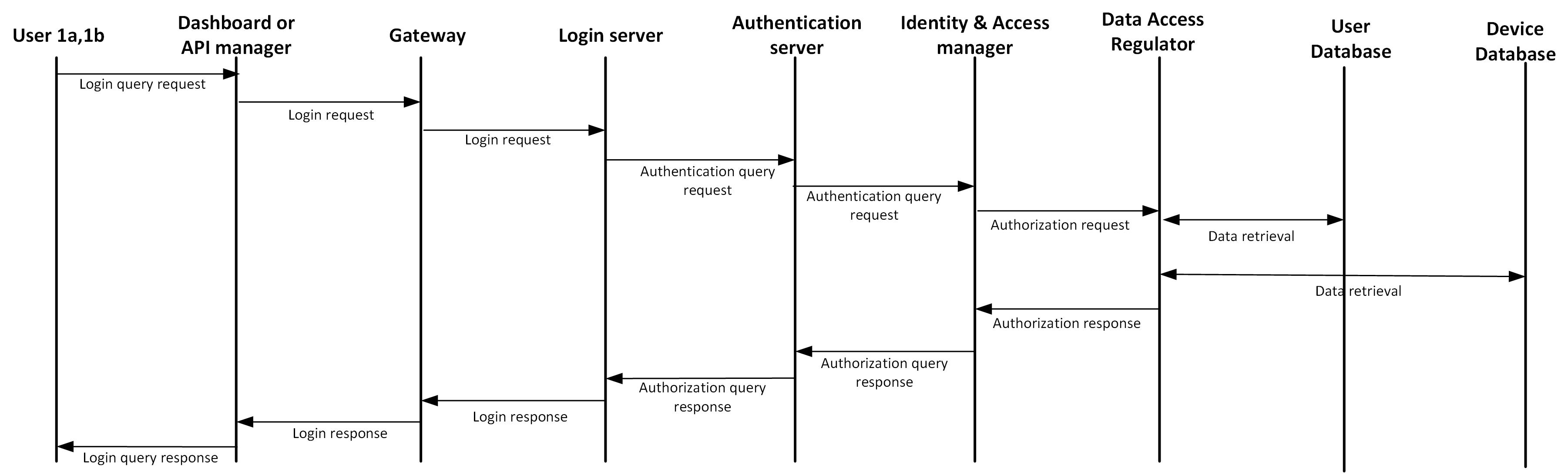}
        \caption{Sequence diagram illustrating the user access management process in a Smart Home (SH) system before data masking implementation in the user database and  end-to-end encryption}
        \label{fig:user acccess mgt-b4}
\end{figure*}

\begin{figure*}[ht!]
        \centering
        \includegraphics[width=17cm,height=8.5cm]{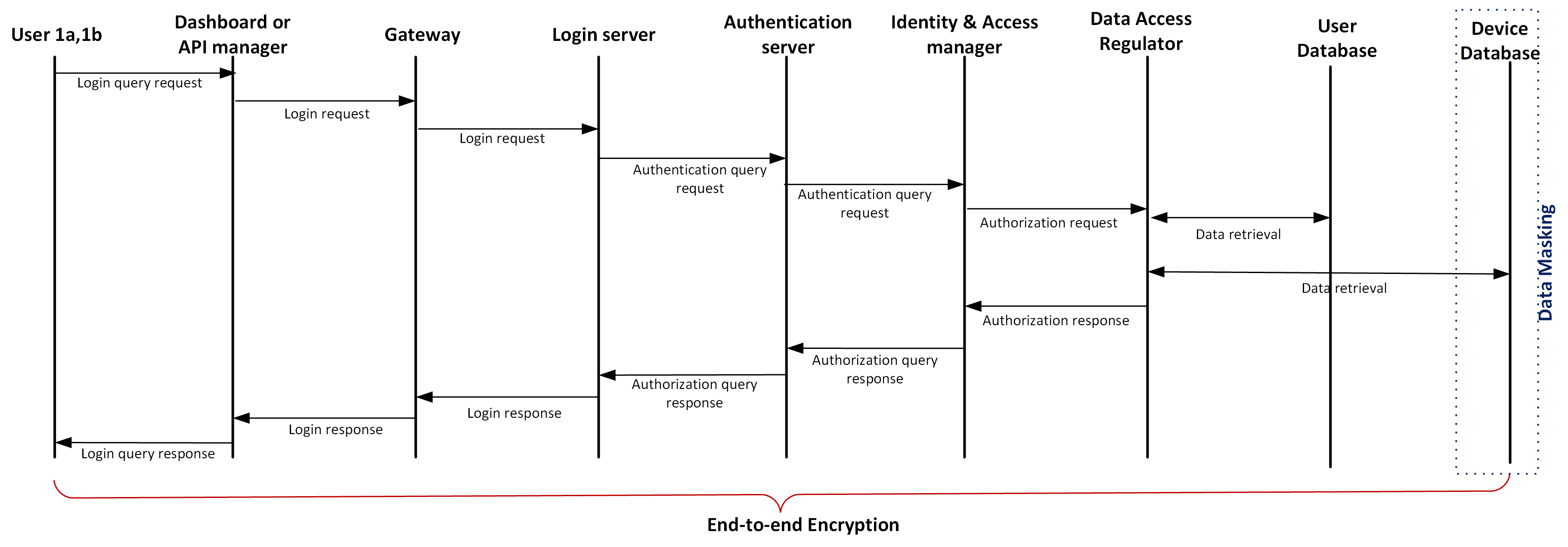}
        \caption{Sequence diagram illustrating the access management process in a Smart Home (SH) system after data masking implementation in the user database and  end-to-end encryption}
        \label{fig:user access mgt after}
\end{figure*}

This analysis primarily focuses on the preservation of user and device data during the user access management process and device commissioning/activity management, as detailed in \autoref{table:map-table}. Data specifically related to user access management process, extracted from \autoref{table:map-table}, is summarized in \autoref{table:map-table1}. Furthermore, the sequence diagrams in \autoref{fig:user acccess mgt-b4} and \autoref{fig:user access mgt after} illustrate the user access management process before and after the implementation of data masking within the user database, respectively.

As indicated in \autoref{table:map-table1}, privacy threats T1, T2, T3, T4, T5, T6, T8, and T11 are identified as having a likelihood of occurrence during user access management. However, the introduction of data masking to protect user data at rest, in conjunction with the deployment of end-to-end encryption mechanisms such as Conditional Proxy Re-Encryption (CPRE) \cite{lin2024end} and JEDI (Joining Encryption and Delegation for IoT) \cite{kumar2019jedi}, will significantly reduce the likelihood of these threats by securing data in transit.

\begin{figure*}[ht!]
        \centering
        \includegraphics[width=16cm,height=7cm]{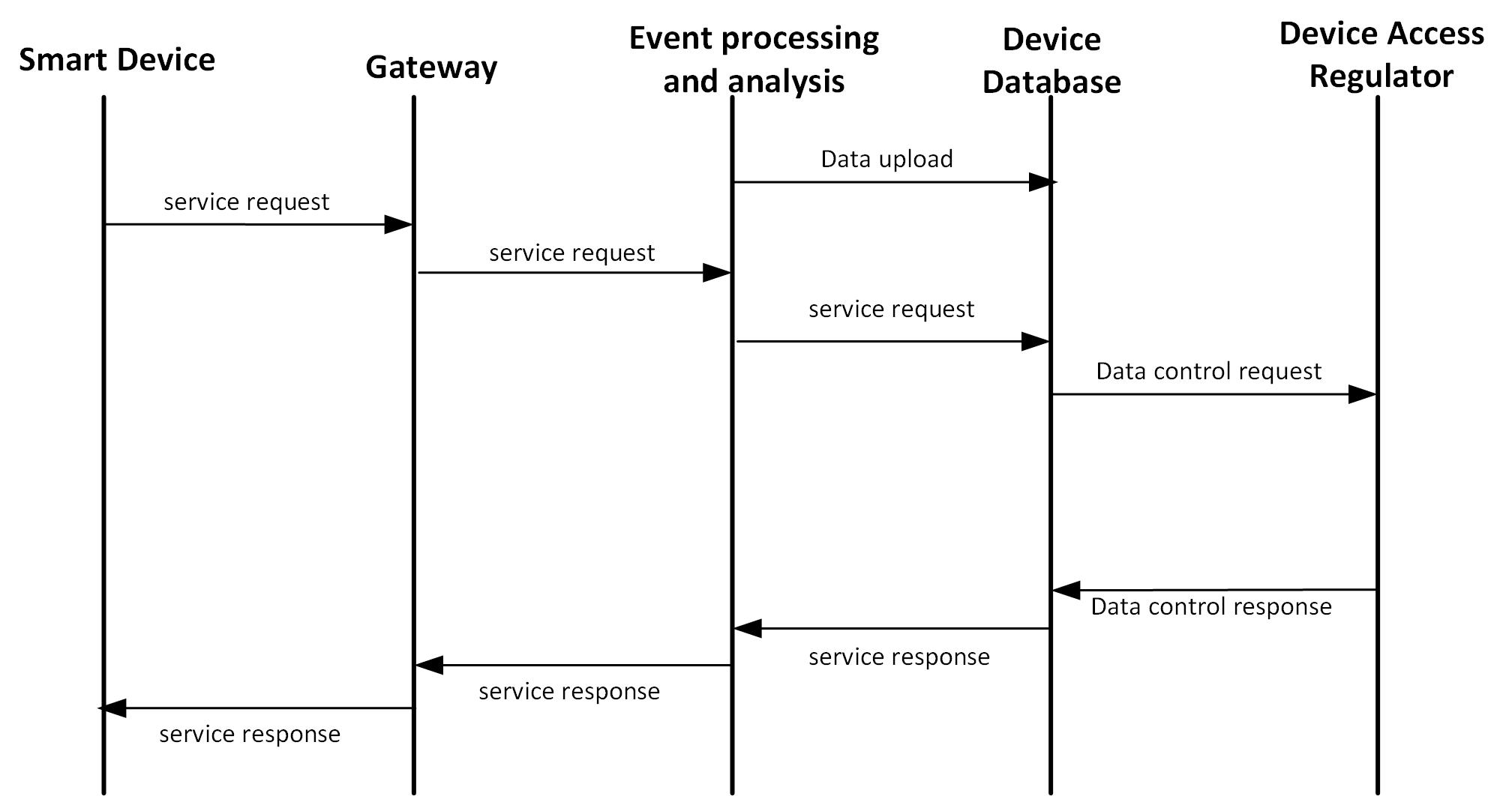}
        \caption{Sequence diagram illustrating the device commissioning process in a Smart Home (SH) system before data masking implementation in the device database and end-to-end encryption}
        \label{fig:Device-comm}
\end{figure*}

\begin{figure*}[ht!]
        \centering
        \includegraphics[width=16cm,height=7cm]{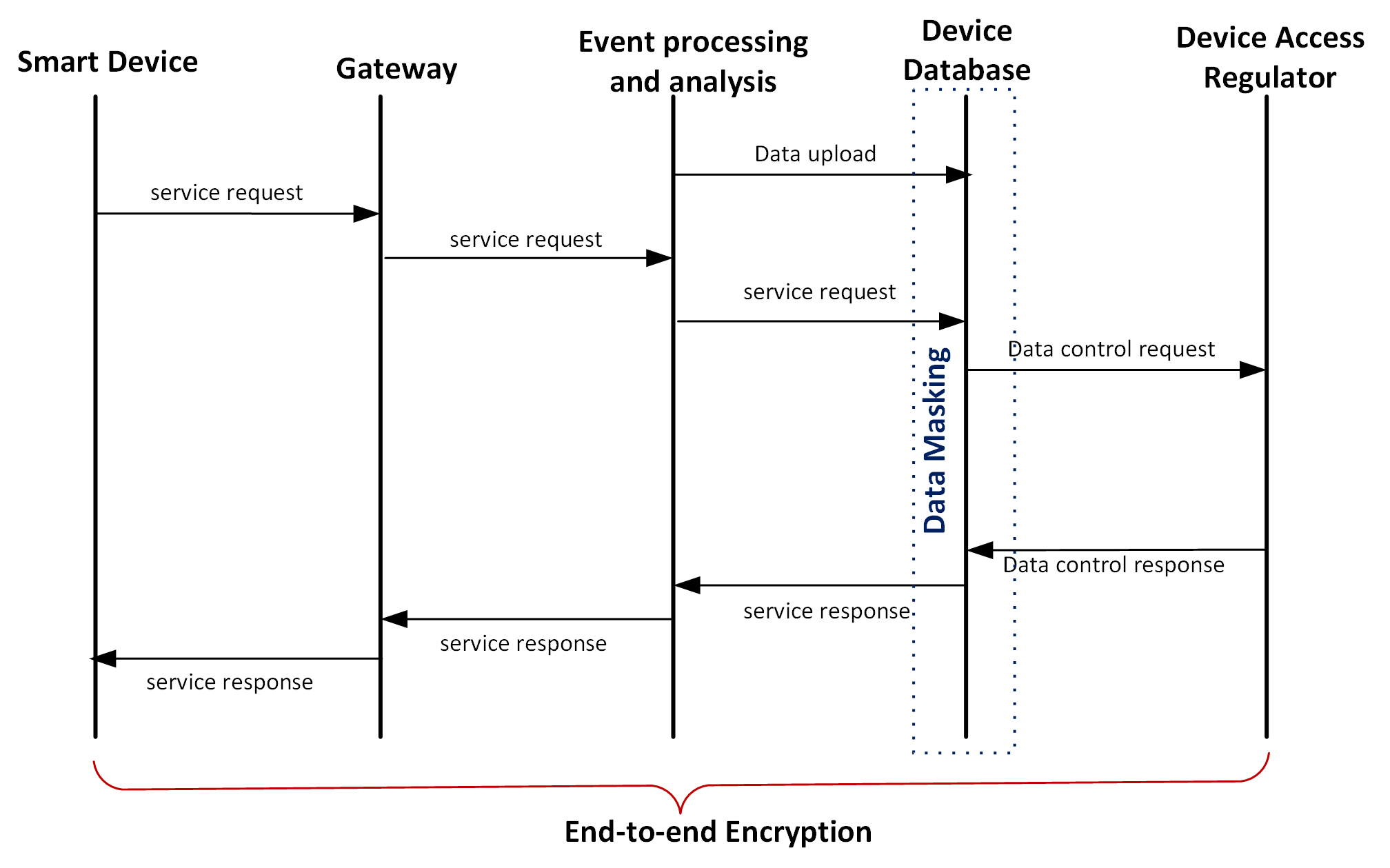}
        \caption{Sequence diagram illustrating the device commissioning process in a Smart Home (SH) system after data masking implementation in the user database and end-to-end encryption}
        \label{fig:Device-Comm after}
\end{figure*}

In contrast, \autoref{table:map-table2} extracted from \autoref{table:map-table} identifies T1, T3, T4, T7, T10, and T11 as privacy threats likely to occur during device commissioning/activity management. The sequence diagrams in \autoref{fig:Device-comm} and \autoref{fig:Device-Comm after} depict the device commissioning and activity management processes before and after the implementation of data masking and end-to-end encryption, respectively. 

The application of data masking within the device database is expected to protect device data at rest. The implementation of end-to-end encryption will further preserve data in transit during interactions among entities involved in device commissioning/activity management. The adoption of these PETs is projected to reduce the likelihood of occurrence of each identified privacy threat to zero. By eliminating all known exploitable vulnerabilities within these processes, these privacy measures substantially reduce the overall privacy risk in the SH system. 

To verify the impact of the assumed PETs introduced, the likelihood of occurrence for each privacy threat is recalculated based on the updated values for user access management and device commissioning/activity management using \autoref{eq:new_table}.

\begin{equation}
    T'_n=T_n-(T_u+T_d)
    \label{eq:new_table}
\end{equation}

Where $T'_n$ is the total number of possible privacy threat occurrences in all interactions after the implementation of data masking in the databases and end-to-end encryption during user access management, and device commissioning/ activity management process, $T_u$ is the total number of possible privacy threat occurrences during user access management and $T_d$ is the total number of possible occurrences of privacy threat during device commissioning/ activity management. $T_n$ is the total number of occurrences of a given privacy threat before the implementation of PET (see \autoref{eq:3}).

It is essential to note that the implementation of PETs in our study primarily affects the likelihood of privacy threats occurring within the specific processes (user access management and device commissioning/activity management) to which these technologies are applied in our case study. However, other processes, like third-party access, may remain vulnerable to these threats. Moreover, the consequences ($C$) associated with privacy threats remain unchanged as a result of PET implementation. This is because the aggravation of a realized privacy threat remains the same, regardless of where it originates in the SH system. In essence, while PETs can effectively reduce the likelihood of privacy threats occurring within targeted processes, the potential severity of their consequences remains system-wide.

Following the implementation of PETs in our case study, the overall likelihood of each privacy threat occurring within the SH system significantly reduced. The results of these privacy measures are summarized in Table \ref{tab:PIA_After_PET}. The key observations from our analysis, conducted after the application of data masking and end-to-end encryption, are as follows:

\begin{itemize}
    \item The risk potential for T1, T3, T4, T9, and T10 remains low; however, their privacy risk value decreases, especially for T1-Identification of SH element. This means that the PETs implementation would eliminate the privacy risk of exposing device identity entirely. 
    \item The risk potential for T2, T5, T7 changes from Moderate to Low with a significant decrease in privacy risk value.
    \item The privacy threats T6 and T8, with an initial high risk potential, are reduced to Moderate.
    \item The privacy threat T11-inventory attack was most impacted by the implementation of data masking and the end-to-end encryption process. T11 risk potential was decreased from high to Low.  
\end{itemize}

\begin{table*}[ht!]
\fontsize{7pt}{7pt}\selectfont
\renewcommand{\arraystretch}{1.3}
\addtolength{\tabcolsep}{-0.8pt}
\begin{longtable}{|>{\raggedright\arraybackslash}p{2.4cm}|>
{\raggedright\arraybackslash}p{2.4cm}|>{\raggedright\arraybackslash}p{2.4cm}|
*{11}{>{\centering\arraybackslash}p{0.1cm}|}}
\caption{An interaction-based mapping of entities involved in User Access Management in the Smart Home (SH) system}
 \label{table:map-table1}\\
\hline
\multicolumn{3}{|c|}{\textbf{Interaction}} & \multicolumn{11}{c|}{\textbf{Privacy Threat}} \\
\hline
\textbf{Source} & \textbf{Flow} & \textbf{Destination} & \textbf{T1} & \textbf{T2} & \textbf{T3} & \textbf{T4} & \textbf{T5} & \textbf{T6} & \textbf{T7} & \textbf{T8} & \textbf{T9} & \textbf{T10} & \textbf{T11} \\
\hline
\endfirsthead

\hline
\multicolumn{3}{|c|}{\textbf{Interaction}} & \multicolumn{11}{c|}{\textbf{Privacy Threat}} \\
\hline
\textbf{Source} & \textbf{Flow} & \textbf{Destination} & \textbf{T1} & \textbf{T2} & \textbf{T3} & \textbf{T4} & \textbf{T5} & \textbf{T6} & \textbf{T7} & \textbf{T8} & \textbf{T9} & \textbf{T10} & \textbf{T11} \\
\hline
\endhead

\hline
\multicolumn{14}{|c|}{\textbf{User access management process}} \\ \hline
\hline

User (1a,1b) & Login query request & Dashboard or API manager (3) &  & x &  & x & x & x &  & x &  &  &  \\ \hline
Dashboard or API manager (3) & Login request & Gateway (12) &  & x &  & x & x & x &  & x &  &  &  \\ \hline
Gateway (12) & Login request & Login server (7) &  & x & x &  &  & x &  & x &  &  & x \\ \hline
Login server (7) & Authentication query request & Authentication server (6) &  &  &  &  &  &  &  &  &  &  &  \\ \hline
Authentication server (6) & Authentication query request & Identity and access manager (9) &  &  &  &  &  &  &  &  &  &  &  \\ \hline
Identity and access manager (9) & Authorization request & Data access regulator (11) &  &  &  &  &  &  &  &  &  &  &  \\ \hline
Data access regulator (11) & Data retrieval & User Database (5a) &  &  &  &  &  &  &  &  &  &  &  \\ \hline
Data access regulator (11) & Data retrieval & Device Database (5b) &  &  &  &  &  &  &  &  &  &  &  \\ \hline
Data access regulator (11) & Authorization response & Identity and access manager (9) &  &  &  &  &  &  &  &  &  &  &  \\ \hline
Identity and access manager (9) & Authorization query response & Authentication server (6) &  &  &  &  &  &  &  &  &  &  &  \\ \hline
Authentication server (6) & Authorization query response & Login server (7) &  &  &  &  &  &  &  &  &  &  &  \\ \hline
Login server (7) & Login response & Gateway (12) &  & x & x &  & x & x &  & x &  &  & x \\ \hline
Gateway (12) & Login response & Dashboard or API manager (3) & x & x & x & x & x & x &  & x &  &  & x \\ \hline
Dashboard or API manager (3) & Login query response & User (1a,1b) &  & x &  &  & x & x &  & x &  &  &  \\ 
\hline

\multicolumn{3}{|c|}{\textbf{Total number of possible privacy threat occurrence during user access management ($T_u$)}} & \textcolor{red}{1} & \textcolor{red}{6} & \textcolor{red}{3} & \textcolor{red}{3} & \textcolor{red}{5} & \textcolor{red}{6} & 0 & \textcolor{red}{6} & 0 & 0 & \textcolor{red}{3} \\
\hline

\end{longtable}
\end{table*}

\begin{table*}[ht!]

\centering
\fontsize{7pt}{8pt}\selectfont
\renewcommand{\arraystretch}{1.3}
\setlength{\tabcolsep}{2pt} 
\begin{longtable}{
    |>{\raggedright\arraybackslash}p{2.4cm}
    |>{\raggedright\arraybackslash}p{2.4cm}
    |>{\raggedright\arraybackslash}p{2.4cm}
    *{11}{|>{\centering\arraybackslash}p{0.45cm}}
    |}
\caption{Interaction-based mapping of entities involved in Device Commissioning/Activity Management in the Smart Home (SH) system}
\label{table:map-table2}\\
\hline
\multicolumn{3}{|c|}{\textbf{Interaction}} & \multicolumn{11}{c|}{\textbf{Privacy Threat}} \\
\hline
\textbf{Source} & \textbf{Flow} & \textbf{Destination} & \textbf{T1} & \textbf{T2} & \textbf{T3} & \textbf{T4} & \textbf{T5} & \textbf{T6} & \textbf{T7} & \textbf{T8} & \textbf{T9} & \textbf{T10} & \textbf{T11} \\
\hline
\endfirsthead

\hline
\multicolumn{3}{|c|}{\textbf{Interaction}} & \multicolumn{11}{c|}{\textbf{Privacy Threat}} \\
\hline
\textbf{Source} & \textbf{Flow} & \textbf{Destination} & \textbf{T1} & \textbf{T2} & \textbf{T3} & \textbf{T4} & \textbf{T5} & \textbf{T6} & \textbf{T7} & \textbf{T8} & \textbf{T9} & \textbf{T10} & \textbf{T11} \\
\hline
\endhead

\hline
\multicolumn{14}{|c|}{\textbf{Device commissioning/activity management process}} \\
\hline
Smart Device (13) & Commissioning request & Gateway (12) & x &   &   &   &   &   & x &   &   &   & x \\ \hline
Gateway (12) & Commissioning request & Event Processing and Analysis (10) & x &   &   &   &   &   & x &   &   &   & x \\ \hline
Event Processing and Analysis (10) & Data upload & Device Database (5b) &   &   &   &   &   &   &   &   &   & x & x \\ \hline
Smart Device (13) & Service request & Gateway (12) & x &   & x &   &   &   & x &   &   &   & x \\ \hline
Gateway (12) & Service request & Event Processing and Analysis (10) & x &   & x & x &   &   & x &   &   &   & x \\ \hline
Event Processing and Analysis (10) & Service request & Device Database (5b) &   &   &   &   &   &   &   &   &   &   &   \\ \hline
Device Database (5b) & Data control request & Data Access Regulator (11) &   &   &   &   &   &   &   &   &   &   &   \\ \hline
Data Access Regulator (11) & Data control response & Device Database (5b) &   &   &   &   &   &   &   &   &   &   &   \\ \hline
Device Database (5b) & Service response & Event Processing and Analysis (10) &   &   &   &   &   &   &   &   &   &   &   \\ \hline
Event Processing and Analysis (10) & Service response & Gateway (12) & x &   & x & x &   &   & x &   &   &   & x \\ \hline
Gateway (12) & Service response & Smart Device (13) & x &   & x &   &   &   & x &   &   &   & x \\ \hline
\hline
\multicolumn{3}{|c|}{\textbf{Total number of possible privacy threat occurrences during device commissioning/activity management ($T_d$)}} & 
\textcolor{red}{6} & 0 & \textcolor{red}{4} & \textcolor{red}{2} & 0 & 0 & \textcolor{red}{6} & 0 & 0 & \textcolor{red}{1} & \textcolor{red}{7} \\
\hline
\end{longtable}

\end{table*}

\begin{table*}[ht!]
\caption{A table showing the prioritization of privacy threats after the implementation of data masking and end-to-end encryption during user access management and device commissioning/activity management}
\label{tab:PIA_After_PET}
\centering
\small
\renewcommand{\arraystretch}{1.15}
\rowcolors{2}{gray!10}{white}
\begin{tabular}{
    c  
    c  
    c  
    c  
    c  
    c  
    c  
    c  
    >{\columncolor{white}}c  
}
\toprule
\textbf{Privacy Threats} & $T_n$ & $T_u$ & $T_d$ & $T'_n=T_n-(T_u+T_d)$ & L = $T'_n$/$T_i$ & C & PIA=L$\times$C & \textbf{Prioritization} \\
\midrule
T1   & 7  & 1 & 6 & 0 & 0 & 2 & 0 $\approx$ 0.00 & \cellcolor{blue!50}Low \\
T2   & 11 & 6 & 0 & 5 & 0.142857143 & 3 & 0.428571429 $\approx$ 0.43 & \cellcolor{blue!50}Low \\ 
T3   & 8  & 3 & 4 & 1 & 0.028571429 & 2 & 0.057142857 $\approx$ 0.06 & \cellcolor{blue!50}Low \\
T4   & 6  & 3 & 2 & 1 & 0.028571429 & 2 & 0.057142857 $\approx$ 0.06 & \cellcolor{blue!50}Low \\
T5   & 6  & 5 & 0 & 1 & 0.028571429 & 3 & 0.085714286 $\approx$ 0.09 & \cellcolor{blue!50}Low \\
T6   & 13 & 6 & 0 & 7 & 0.200000000 & 3 & 0.600000000 $\approx$ 0.60 & \cellcolor{yellow!100}Moderate \\
T7   & 6  & 0 & 6 & 0 & 0.000000000 & 3 & 0.000000000 $\approx$ 0.00  & \cellcolor{blue!50}Low \\
T8   & 11 & 6 & 0 & 5 & 0.142857143 & 5 & 0.714285714 $\approx$ 0.71 & \cellcolor{yellow!100}Moderate \\
T9   & 1  & 0 & 0 & 1 & 0.028571429 & 1 & 0.028571429 $\approx$ 0.03 & \cellcolor{blue!50}Low \\
T10  & 2  & 0 & 1 & 1 & 0.028571429 & 3 & 0.085714286 $\approx$ 0.09 & \cellcolor{blue!50}Low \\
T11  & 13 & 3 & 7 & 3 & 0.085714286 & 5 & 0.428571429 $\approx$ 0.43 & \cellcolor{blue!50}Low \\
\bottomrule
\end{tabular}

\end{table*}

\subsection{Alternative Methods for Mitigating Privacy Threats}

In addition to the data masking technique employed in this study, several other PETs can be utilized to ensure the privacy-preserved storage of user and device data. Examples of some of these PETs include \textbf{Differential Privacy (DP)} \cite{waheed2023privacy}, and \textbf{Pseudonymization} \cite{al2019misty,suomalainen2016enhancing}. While the implementation of these PETs typically requires substantial computational resources, de-identification methods such as \textbf{Selective Differential Privacy (SDP)} offer a more resource-efficient alternative \cite{shi-etal-2022-selective}. Selective DP operates similarly to Standard Differential Privacy (DP) but introduces controlled noise tailored explicitly to the constraints of SH systems, making SDP a preferable option over DP due to limitations in computational capacity and device heterogeneity in the SH system.

As an alternative to the end-to-end encryption approach discussed in our case study, \textbf{Secure Multiparty Computation} \cite{shin2017secure, xu2019edge} can be employed to protect interactions between privacy service providers and third parties. This method effectively mitigates privacy threats such as T4, T5, T9, and T11, which are associated with third-party access processes.

Furthermore, non-technical measures rooted in policy and regulatory frameworks can significantly contribute to the reduction of privacy risk associated with privacy threats, particularly T4 and T9. The enforcement of such regulations compels service providers and third parties to comply with legal requirements and to implement appropriate technical measures for privacy protection. Examples of these regulatory mitigation include obtaining informed consent, practicing data minimization, and granting data subjects control over their personal data usage, all in accordance with privacy regulations such as the General Data Protection Regulation (GDPR).

\section{Discussion}
\label{section:Discussion}
The integration of PE in our study, as guided by the LINDDUN PRO PE framework illustrated in \autoref{fig:LINDDUN-framework}, provides a systematic approach to implement privacy measures within SH systems. Starting with the presentation of a comprehensive reference model for a typical SH system in \autoref{fig:reference-model}, we develop the DFD in \autoref{fig:DFD}, which illustrates the detailed connections between SH entities and their interactions within the SH system. These two processes represent the first stage of the PE framework, denoted as "Model the system" in \autoref{fig:LINDDUN-framework}. The second stage of the LINDDUN PRO PE framework, which is the "Elicit threats," leverages the granularity of the DFD in the first stage to facilitate the elicitation process for identifying potential interactions susceptible to privacy threats within the SH system. Our study identified 11 distinct privacy threats associated with SH systems, as shown in \autoref{table:Summarize-tab}. This interaction-based elicitation process, constituting our study's Privacy Threat Analysis (PTA), involves a detailed assessment of the likelihood of occurrence for each privacy threat, as shown in \autoref{table:map-table}. The final stage of the PE framework involves the strategic Privacy Impact Assessment (PIA) and prioritization of privacy threats based on the calculated privacy risk value of each privacy threat. 

To assess the reliability and applicability of our proposed PE approach within the SH system, we conducted a case study focusing on the preservation of data at rest in both user and device databases through data masking. Additionally, we assumed the implementation of end-to-end encryption during user access management and device commissioning/activity management processes. These PETs were selected and applied to mitigate the likelihood of the occurrence of possible privacy threats within the SH system. The effectiveness of these privacy threat mitigation strategies is summarized in \autoref{tab:PIA_Before_PET} (PIA before PET deployment) and \autoref{tab:PIA_After_PET} (PIA after PET deployment), which illustrate the practical implications of PET deployment in the SH system. Specifically, the tables highlight changes in the likelihood of privacy threat occurrence, the results of the Privacy Impact Assessment (PIA), and the prioritization of threats before and after PET implementation. Importantly, our approach integrates both user and device data within the PE framework, supporting a comprehensive and holistic privacy management strategy for SH systems. Furthermore, our analysis incorporates real-world privacy threats identified from existing literature, enhancing the relevance and applicability of our findings to actual SH systems. In addition to the implemented PETs, we suggest other PETs that could offer additional risk reduction and mitigation of privacy threats in future deployments.

The interaction-based (source-flow-destination) analysis proposed in this study is applicable to any SH systems, regardless of their inherent heterogeneity. This applicability is due to the method’s emphasis on analyzing privacy threats at the interaction level, which allows for the accommodation of diverse smart devices within SH systems. Additionally, the approach facilitates the early identification of privacy threats and associated risks. 

Furthermore, the methodology outlined in this work can also be adapted to other Internet of Things (IoT) applications, provided that the underlying IoT architecture is well-defined and represented through a Data Flow Diagram (DFD). Employing DFDs enables a thorough examination of the specific interactions relevant to the IoT application in question, thereby enhancing the comprehensiveness of privacy threat analysis. It is essential to acknowledge that the consequences, likelihood of occurrence, and overall privacy risks associated with IoT applications can vary significantly. These variations are influenced by factors such as the specific nature of the assets to be protected, the architectural design of the IoT system, and the baseline security measures in place. 

Moreover, the findings of this study highlight important implications regarding the deployment of PETs within Smart Home (SH) systems. As noted by Alwedaei et al. \cite{Alwedaei}, the adoption of PETs often entails significant trade-offs. Specifically, these technologies typically require substantial computational resources, which can constrain their practical implementation and potentially degrade the overall user experience in the SH system. Furthermore, the integration of PETs may necessitate more complex system controls, which could overwhelm users and act as a barrier to widespread adoption. In addition to these technical considerations, non-technical approaches such as comprehensive consent mechanisms and detailed privacy disclosures also present challenges. For instance, while these measures are designed to enhance user autonomy and transparency, they may inadvertently contribute to consent fatigue, resulting in user disengagement. Taken together, these factors underscore the need for a balanced approach that carefully weighs the benefits of privacy preservation against the potential impacts on usability and user acceptance in SH systems.

Future research should focus on the development of lightweight PETs that are specifically designed for resource-constrained systems. In addition, it is essential to devise robust privacy measures that maintain a balance between strong privacy protection and user experience. These measures should not compromise usability or lead to user dissatisfaction due to reduced flexibility and convenience in SH systems usage resulting from the implementation of PETs.

\section{Conclusion}
\label{section:conclusion}
Data privacy measures in any form need to comply with standard privacy regulations and laws to limit the occurrence of privacy threats and their impact.

Our findings indicate that privacy threats in the SH can affect users or smart devices, leading to distrust in the adoption and use of the SH system. Establishing robust privacy preservation mechanisms is therefore essential for fostering user confidence. However, a thorough understanding of the architecture of a typical SH is a foundation for building a robust privacy preservation mechanism. In addition, our analysis highlights the importance of device identity privacy, as smart devices within SH systems frequently store, process, and transmit both user and device data. Compromise of this data may have adverse consequences for both device and user privacy.

To address these challenges, we proposed a comprehensive PE approach for SH systems. This methodology integrates PTA and PIA within the LINDDUN PRO PE framework, employing DFD to systematically represent and analyze privacy threats in the SH system. The applicability of our proposed PE was validated through a case study, in which data masking was implemented in both user and device databases, and end-to-end encryption was applied to critical interactions (user access management and device commissioning/activity management processes) under assumed conditions. The results indicate a substantial reduction in privacy risk across the SH system following the deployment of these PETs.

Furthermore, we prioritized and assessed privacy threats before and after the implementation of PETs. The analysis revealed that, before the application of PETs, privacy threats targeting users posed a greater risk to the SH system compared to those affecting smart devices or third parties. However, since smart devices often store and process user data, threats to device privacy can ultimately impact users as well. The implementation of PETs in our case study effectively mitigated these risks, illustrating that appropriate PETs can significantly reduce the likelihood of privacy threats occurring. Nevertheless, due to resource constraints inherent in SH systems, not all PETs suitable for other IoT applications are directly applicable. Consequently, future research should focus on the identification and development of lightweight PETs designed explicitly for SH systems to ensure adequate privacy threat mitigation.

Finally, while the proposed PE approach is adaptable to various SH systems, differences in system architecture may necessitate modifications to the reference model and DFD. Nonetheless, the core principles of our PE methodology can apply to diverse SH systems and IoT applications, regardless of their specific architecture and configurations.

\section*{Acknowledgement}
This work was partially supported by the Natural Sciences
and Engineering Research Council of Canada (NSERC)
through the NSERC Discovery Grant program.

\section* {Appendix}
\subsection*{List of Common Abbreviations and Their Full Terms }

\begin{table}[ht!]
\fontsize{9pt}{9pt}\selectfont
\caption{List of Abbreviations}
\centering
\begin{tabularx}{\textwidth}{@{} l X @{}}
\toprule
\textbf{Abbreviation} & \textbf{Full Term} \\
\midrule
API    & Application Programming Interface \\
CAV    & Connected and Autonomous Vehicles \\
DFD    & Data Flow Diagram \\
DP     & Differential Privacy \\
FIPS   & Federal Information Processing Standards \\
GPS    & Global Positioning System \\
IAPP   & International Association of Privacy Professionals \\
ICT    & Information and Communication Technology \\
IoT    & Internet of Things \\
JEDI   & Joining Encryption and Delegation for IoT \\
LINDDUN & Linking, Identifying, Non-repudiation, Detecting, Data Disclosure, Unawareness, Non-compliance \\
NIST   & National Institute of Standards and Technology \\
OWASP  & Open Web Application Security Project \\
PbD    & Privacy by Design \\
PE     & Privacy Engineering \\
PET    & Privacy-Enhancing Technologies \\
PHI    & Personal Health Information \\
PIA    & Privacy Impact Assessment \\
PII    & Personally Identifiable Information \\
PTA    & Privacy Threat Analysis \\
SDP    & Selective Differential Privacy \\
SH     & Smart Home \\
SLAAC  & Stateless Address Autoconfiguration \\
SP     & Service Provider \\
STRIDE & Spoofing, Tampering, Repudiation, Information Disclosure, Denial of Service, Elevation of Privilege \\
\bottomrule
\end{tabularx}
\end{table}

 \bibliographystyle{elsarticle-num} 
 \bibliography{reference}

\end{document}